\documentclass[%
reprint,
amsmath,amssymb,
aps,
prb,
floatfix,
]{revtex4-1}

\usepackage{graphicx}
\usepackage{dcolumn}
\usepackage{bm}
\usepackage{caption}
\usepackage{subcaption}
\usepackage{amssymb,amsmath}
\usepackage{gensymb}
\usepackage{url}
\usepackage{float,color}
\usepackage{hyperref}    
\begin{document}
	
	\title{Phonons in Twisted Transition Metal Dichalcogenide Bilayers (``twistnonics"): Ultra-soft Phasons, and a transition from Superlubric to  Pinned Phase}
	
	\author{Indrajit Maity, Mit H. Naik, Prabal K Maiti, H. R. Krishnamurthy }
	\author{Manish Jain}
	\email{mjain@iisc.ac.in}
	\affiliation{
		Centre for Condensed Matter Theory, Department of Physics, \\ Indian Institute of Science, Bangalore-560012
	}%

	\begin{abstract}
		The tunability of the interlayer coupling by twisting one layer with respect to another layer of two dimensional materials provides a unique way to manipulate the phonons and related properties. We refer to this engineering of phononic properties as ``twistnonics". We study the effects of twisting on low frequency shear (SM) and layer breathing (LBM) modes in transition metal dichalcogenide (TMD) bilayer using atomistic classical simulations. We show that these low frequency modes are extremely sensitive to twist and can be used to infer the twist angle. We find unique ``ultra-soft" phason modes (frequency $\lesssim 1\  \mathrm{cm^{-1}}$, comparable to acoustic modes) for any non-zero twist, corresponding to an \textit{effective} translation of the moir{\'e} lattice by relative displacement of the constituent layers in a non-trivial way. Unlike the acoustic modes, the velocity of the phason modes are quite sensitive to twist angle. Also, new high-frequency SMs appear, identical to those in stable bilayer TMD ($\theta = 0\degree/60\degree$), due to the overwhelming growth of stable stacking regions in relaxed twisted structures. Our study reveals the possibility of an intriguing $\theta$ dependent superlubric to pinning behavior and of the existence of ultra-soft modes in \textit{all} two-dimensional (2D) materials.

	\end{abstract}
	
	\maketitle
	
\section{\label{sec:level1}Introduction}
Twisting one layer of a bilayer system with respect to another provides a unique degree of freedom for tuning the properties of two dimensional (2D) materials. For example, in case of bilayer graphene, twisting leads to (a) structural changes, such as the observation of topological point defects, domain walls and layer buckling \cite{Alden_PNAS,yoo_natmat_2019, Sunku_science_2018,moire_with_twist,Zhang_2018_jmps,Oleg_2dm_2018,Huang_topology_prl_2018,Lebedeva_commincomm_prb_2019}, (b) significant change in electronic properties including superconductivity at ``magic" twist angles \cite{Nam_prb_2017, Angeli_arxiv_2019, Bistritzer_pnas_2011, Choi_prbrap_2018, Tarnopolsky_prl_2019, Cao_nature_2018, Cao_strange_2019, Cao_2018_nature,Carr_prb_2017,Kerelsky_arxiv_2018, Lucignano_prb_2019,Alexander_arxiv_2018,Xiaobo_arxiv_2019}, 
and (c) superlubricity, a state of ultra-low friction 
\cite{Dien_2004_PRL, Oded_prb_2012, Wang_prb_2019}. An important facet of twisting is the evolution of low frequency vibrational modes, which has largely remained unexplored. Since the low frequency modes are solely determined by interlayer coupling and are accessible in Raman measurements, they provide a direct, non-destructive probe of the interlayer interaction \cite{Maity_2018_LBM, Zhao_2013_nano, Huang_2016_Nanolet, Puretzky_2016_acsnano}. The existing theoretical reports on the evolution of vibrational modes in twisted structures are restricted to large twist angle and use the Lennard-Jones potential \cite{Cocemasov_prb_2013, Song_physlet_2019} to describe the interlayer interaction, which is insufficient for capturing the stacking dependent energetics \cite{mit_kc_2018}. Although existing experimental studies\cite{Huang_2016_Nanolet, Puretzky_2016_acsnano, Lin_acsnano_2018} have explored small twist angles, they can only probe Raman active modes with frequencies $>10\ \mathrm{cm^{-1}}$.

In this work, we computationally investigate the effects of twisting on low frequency shear (SM) and layer breathing (LBM) modes in bilayer $\mathrm{MoS_{2}}$, a prototypical transition metal dichalcogenide (TMD). Relative in-plane and out-of-plane displacements of the constituent layers give rise to SM and LBM, respectively. The coexistence of several stackings in the moir{\'e} superlattice (MSL) that results from the twist leads to inhomogeneous interlayer coupling. As a consequence, the low frequency modes mix and become quite sensitive to twist. Our calculations show the existence of ultra-soft phason modes, large variation in LBM frequencies, appearance of multiple LBMs and high frequency SM in twisted bilayer $\mathrm{MoS_{2}}$ ($\mathrm{tBLMoS_{2}}$). Moreover, we find the velocity of the phason modes are quite sensitive to twist angle. These observations are generic to TMDs and we confirm our results for $\mathrm{MoSe_{2}}$ as well. The domain walls and point defects present in twisted structures that are inevitable consequences of structural relaxation, are likely to influence the electronic properties \cite{mit_twister, Fleischmann_arxiv_2019, Fengcheng_prl_2019,Liheng_supconduct_arxiv_2019}.
	
The paper is organized in the following manner : In Sec.~II, we detail
the methods used to perform molecular dynamics simulations and computation of phonon frequencies. In Sec.~III A we show the effects of relaxation on the rigidly twisted structures using interlayer separation landscape. In Sec.~III B we discuss mode mixing due to the existence of multiple coexistent stackings. In Sec.~III C we show the twist angle dependence of shear and layer breathing modes. In Sec.~III D we discuss the origin of phason modes in twisted structures and it's role in determining frictional properties of the system. In Sec.~IV we show twist angle dependence of phonon frequencies of twisted bilayer $\mathrm{MoSe_{2}}$ and discuss experiments that can be used to test our predictions. We also discuss the effects of manipulation of phonons on other properties such as specific heat etc. In Sec.~V we summarize our results.

	\begin{figure*}[!htbp]
		\begin{subfigure}[H!]{0.45\textwidth}
			\caption{$\theta=2.9\degree$}
			\label{fig:fig1a}
			\includegraphics[width=\textwidth]{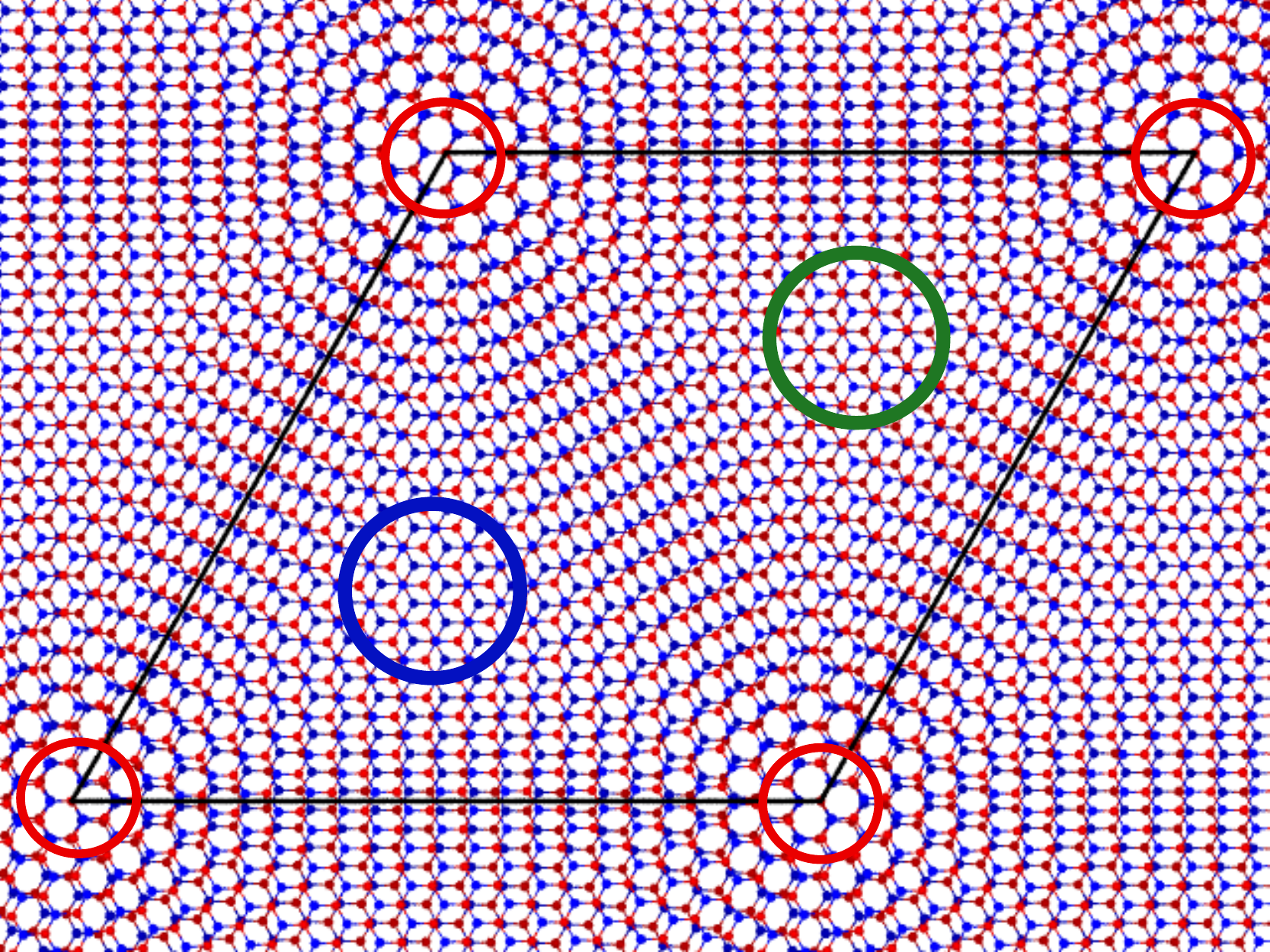}  
		\end{subfigure}
		\hspace{0.5cm}
		\begin{subfigure}[H!]{0.45\textwidth}
			\caption{$\theta=57.1\degree$}
			\label{fig:fig1b}
			\includegraphics[width=\textwidth]{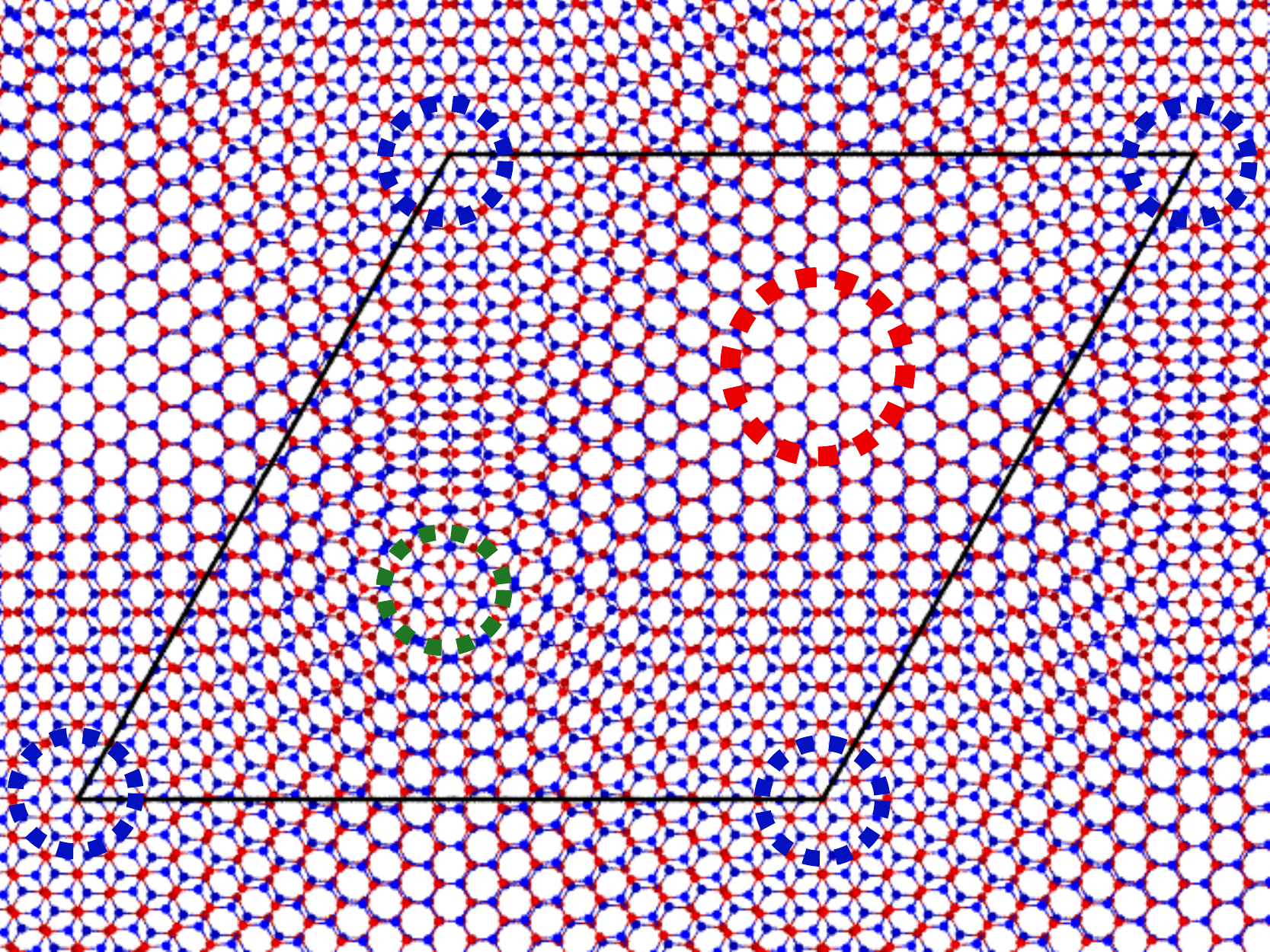} 
		\end{subfigure}
		
		\vspace{0.2cm}
		
		\begin{subfigure}[H!]{0.12\textwidth}
			\caption{AA}
			\label{fig:fig1c}
			\includegraphics[width=\textwidth]{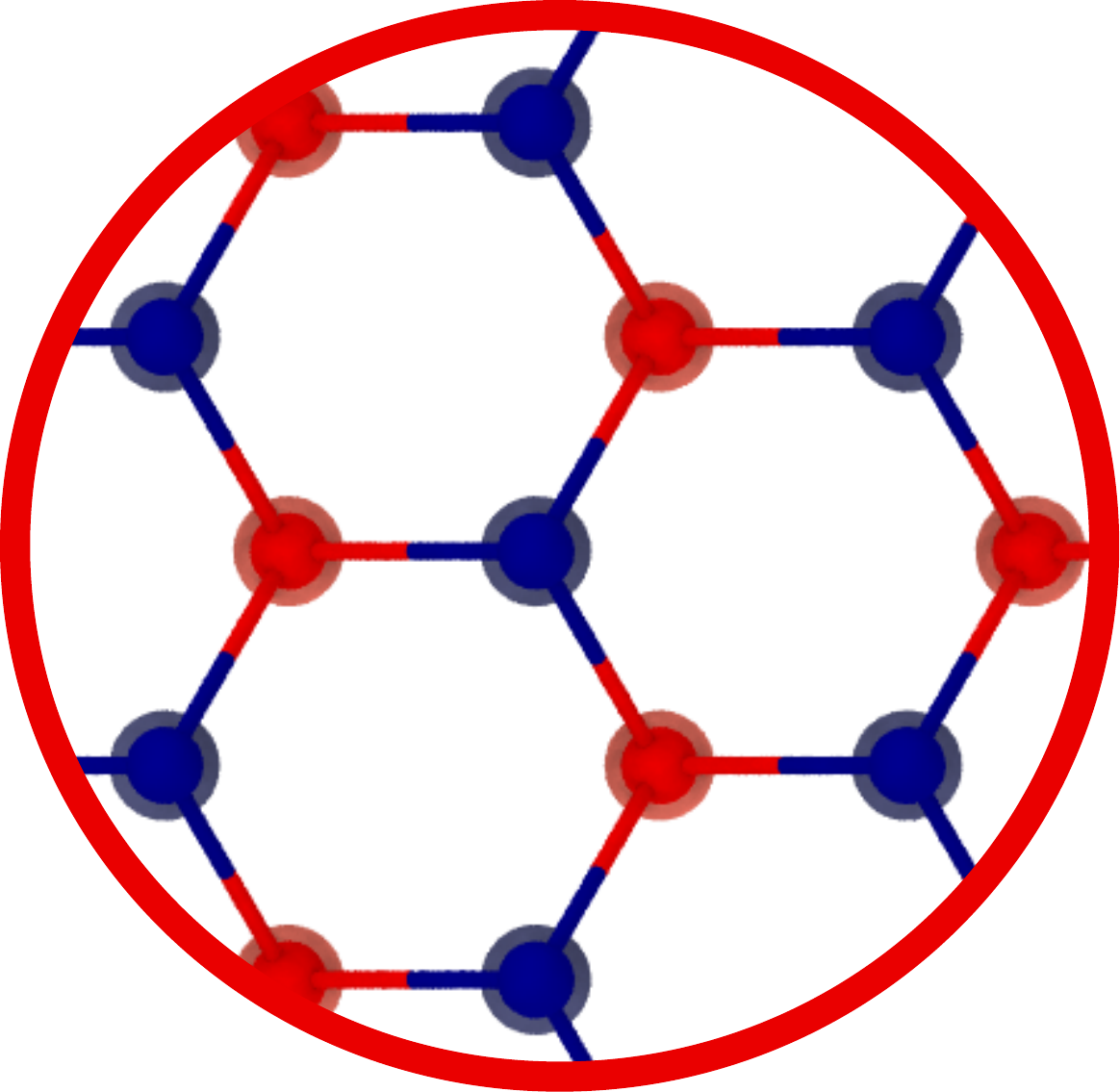}
		\end{subfigure}
		\hspace{0.4cm}
		\begin{subfigure}[H!]{0.12\textwidth}
			\caption{$\mathrm{AB}$}
			\label{fig:fig1d}
			\includegraphics[width=\textwidth]{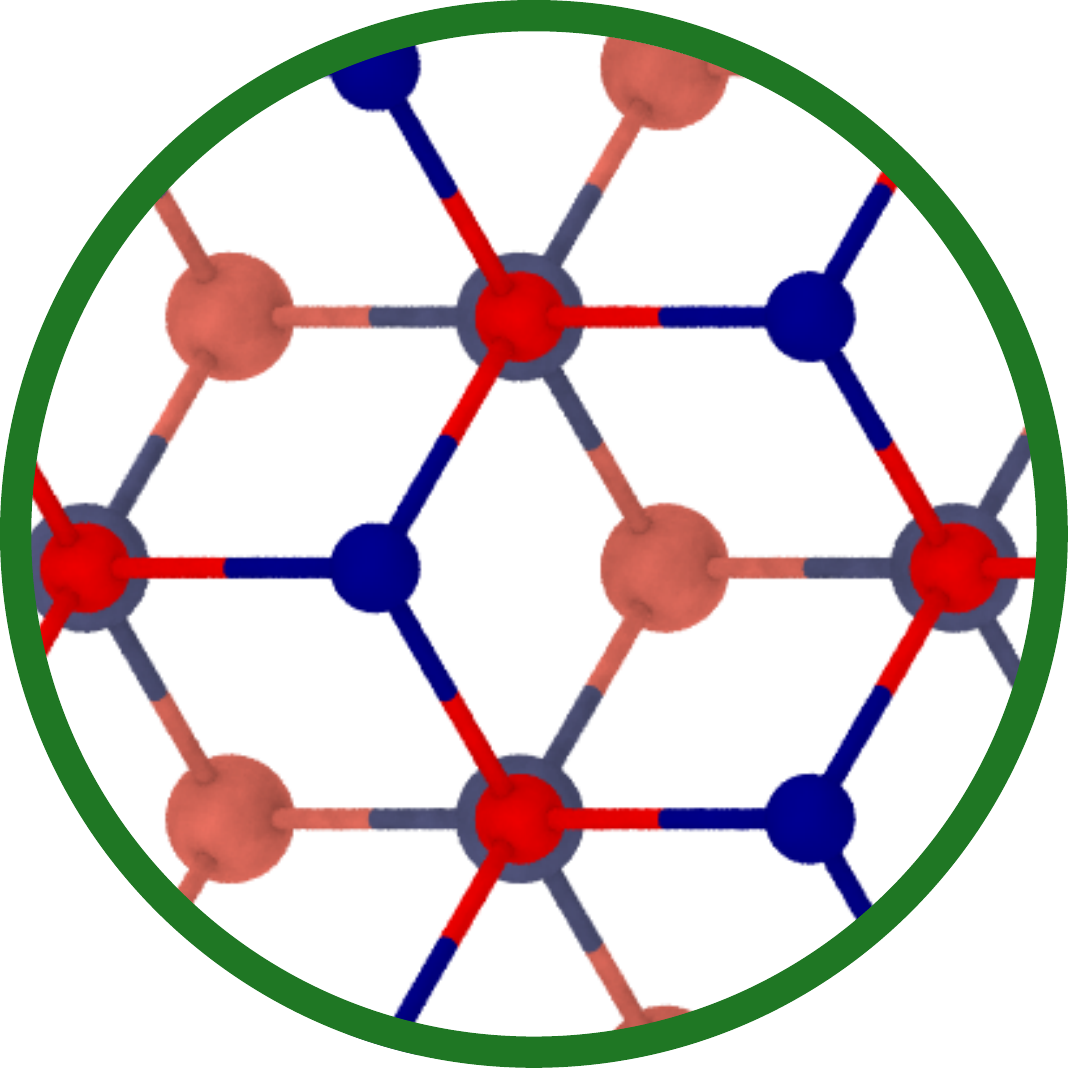}
		\end{subfigure}
		\hspace{0.4cm}
		\begin{subfigure}[H!]{0.12\textwidth}
			\caption{$\mathrm{BA}$}
			\label{fig:fig1e}
			\includegraphics[width=\textwidth]{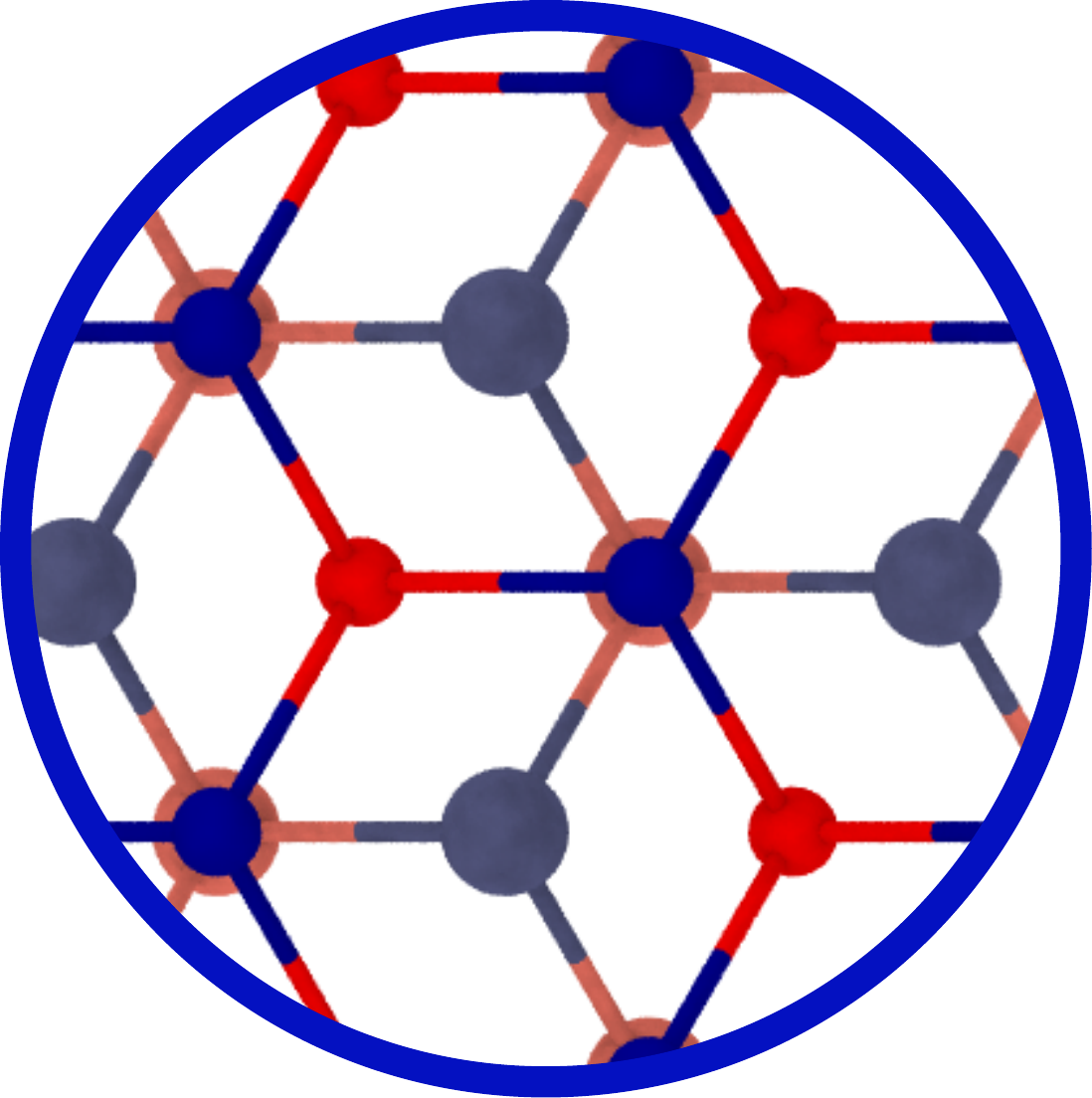}
		\end{subfigure}
		\hspace{0.6cm}
		\begin{subfigure}[H!]{0.12\textwidth}
			\caption{$\mathrm{AA^\prime}$}
			\label{fig:fig1f}
			\includegraphics[width=\textwidth]{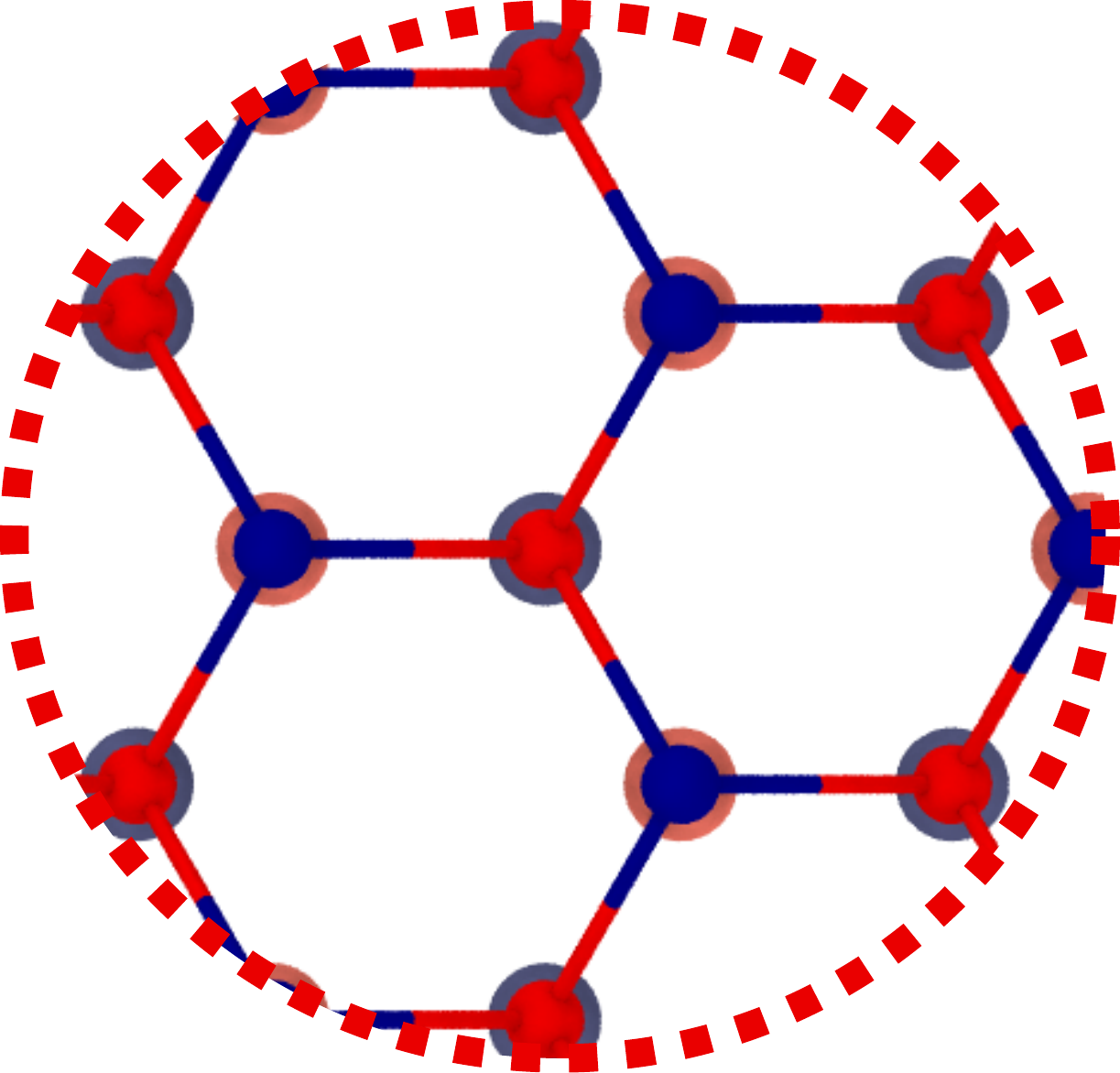}
		\end{subfigure}
		\hspace{0.2cm}
		\begin{subfigure}[H!]{0.16\textwidth}
			\caption{$\mathrm{AB^\prime}$}
			\label{fig:fig1g}
			\includegraphics[width=\textwidth]{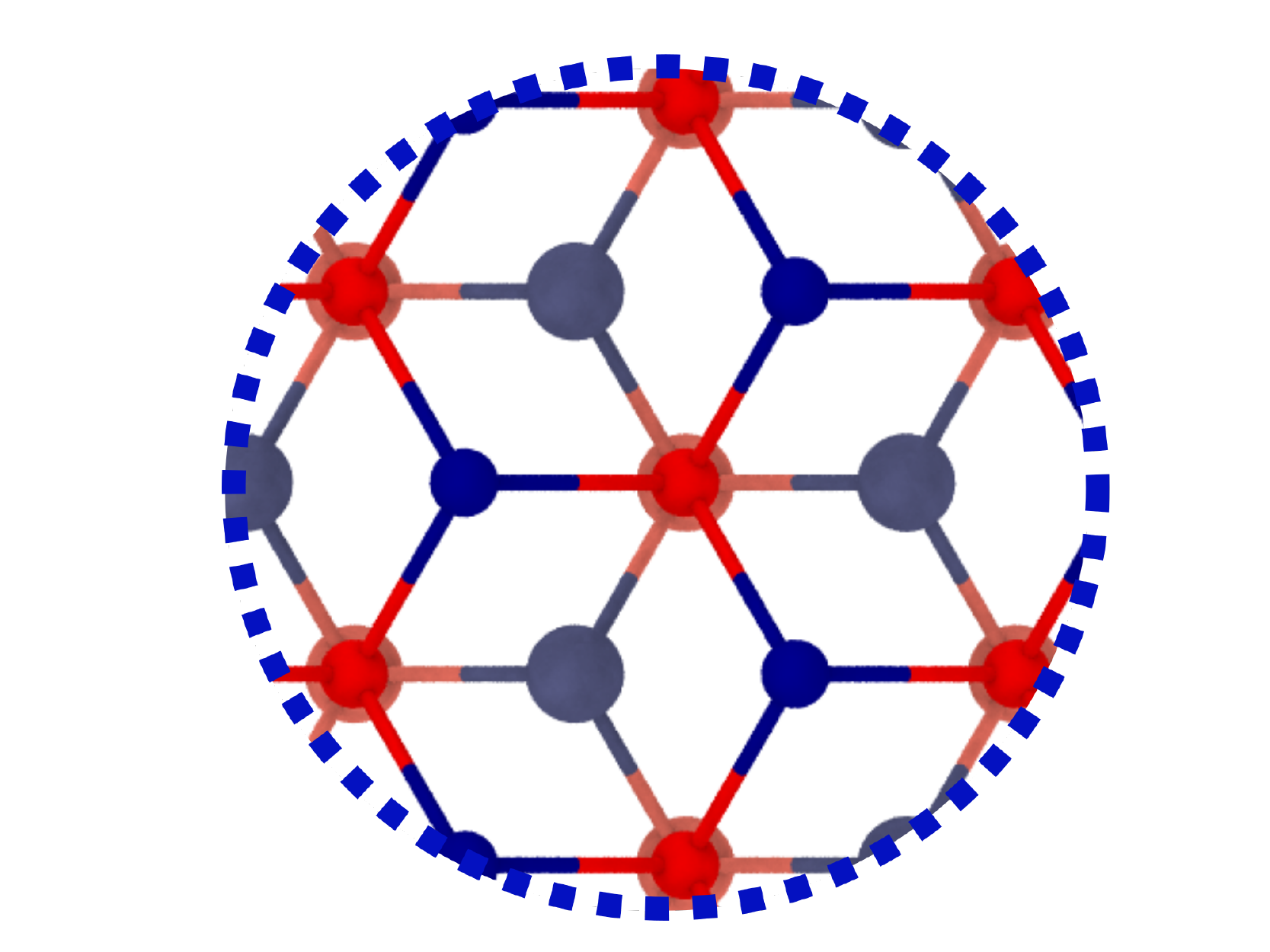}
		\end{subfigure}
		\hspace{0.2cm}
		\begin{subfigure}[H!]{0.115\textwidth}
			\caption{$\mathrm{A^\prime B}$}
			\label{fig:fig1h}
			\includegraphics[width=\textwidth]{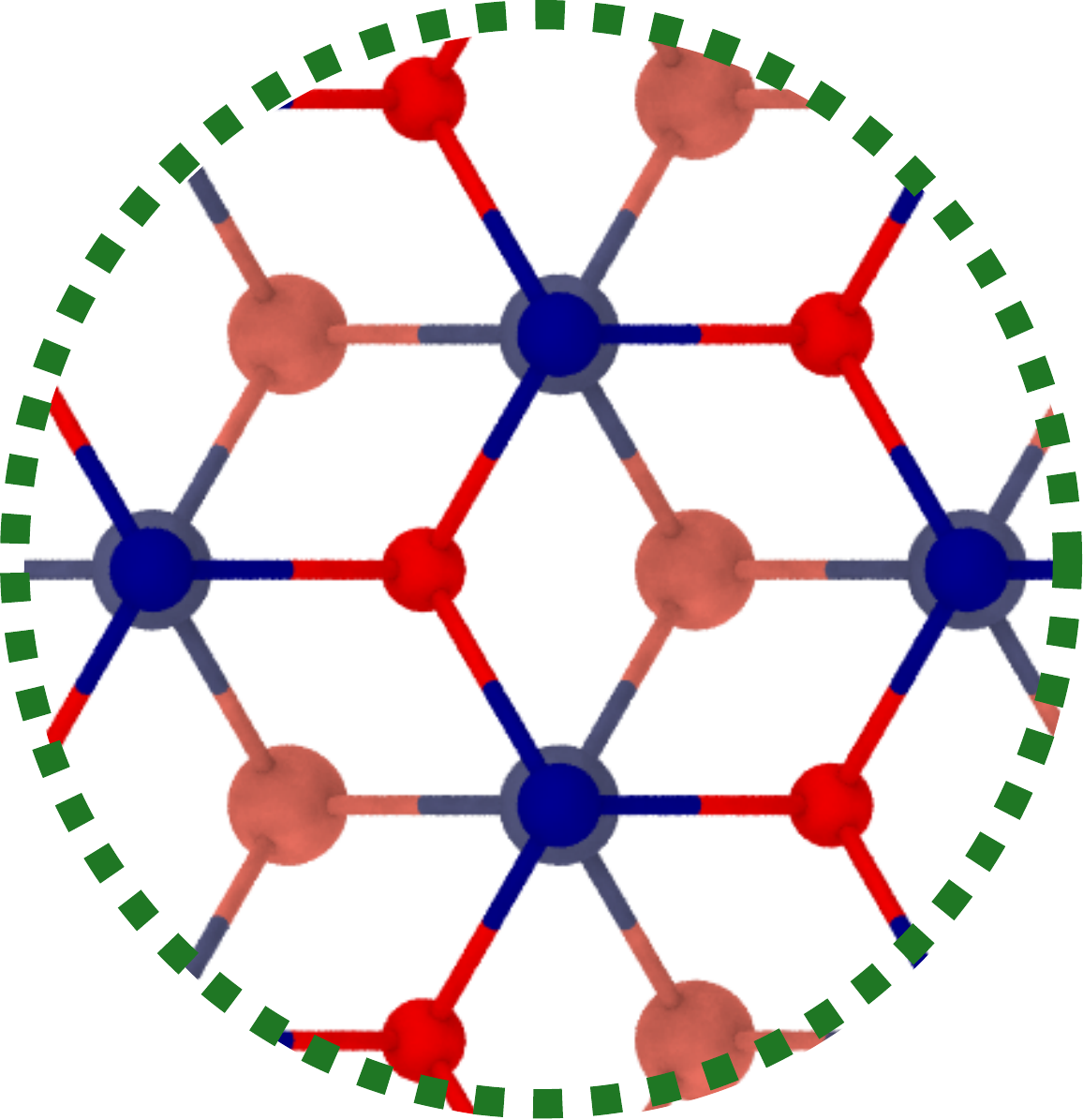}
		\end{subfigure}
		\caption{Relaxed twisted bilayer of $\mathrm{MoS_{2}}$ and the coexistence of multiple high-symmetry stacking regions. For $\theta=2.9\degree$ ($57.1\degree$), high-symmetry stacking regions are marked with solid (dashed) circles. Mo atoms of bottom (top) layer are depicted as large (small) in size and with faded (dark) red color. Similary, we use blue color for S atoms.}
		\label{fig:fig1}
	\end{figure*}	
	
	\begin{figure}[!htbp]
		\begin{subfigure}[t]{0.5\textwidth}
			\vspace*{-0.2cm}
			\includegraphics[width=\textwidth]{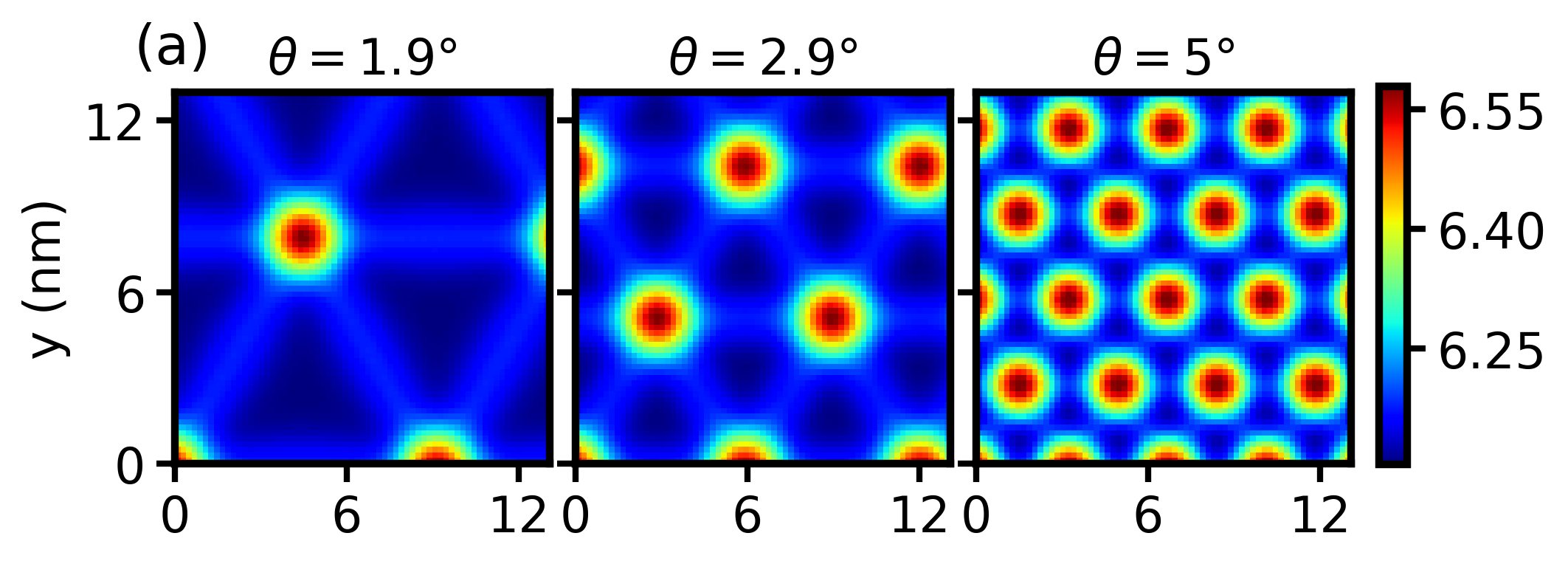}
			\captionsetup{labelformat=empty}
			\caption{}
			\label{fig:rel_ils_a}
		\end{subfigure}
		\begin{subfigure}[t]{0.5\textwidth}
			\vspace*{-0.9cm}
			\includegraphics[width=\textwidth]{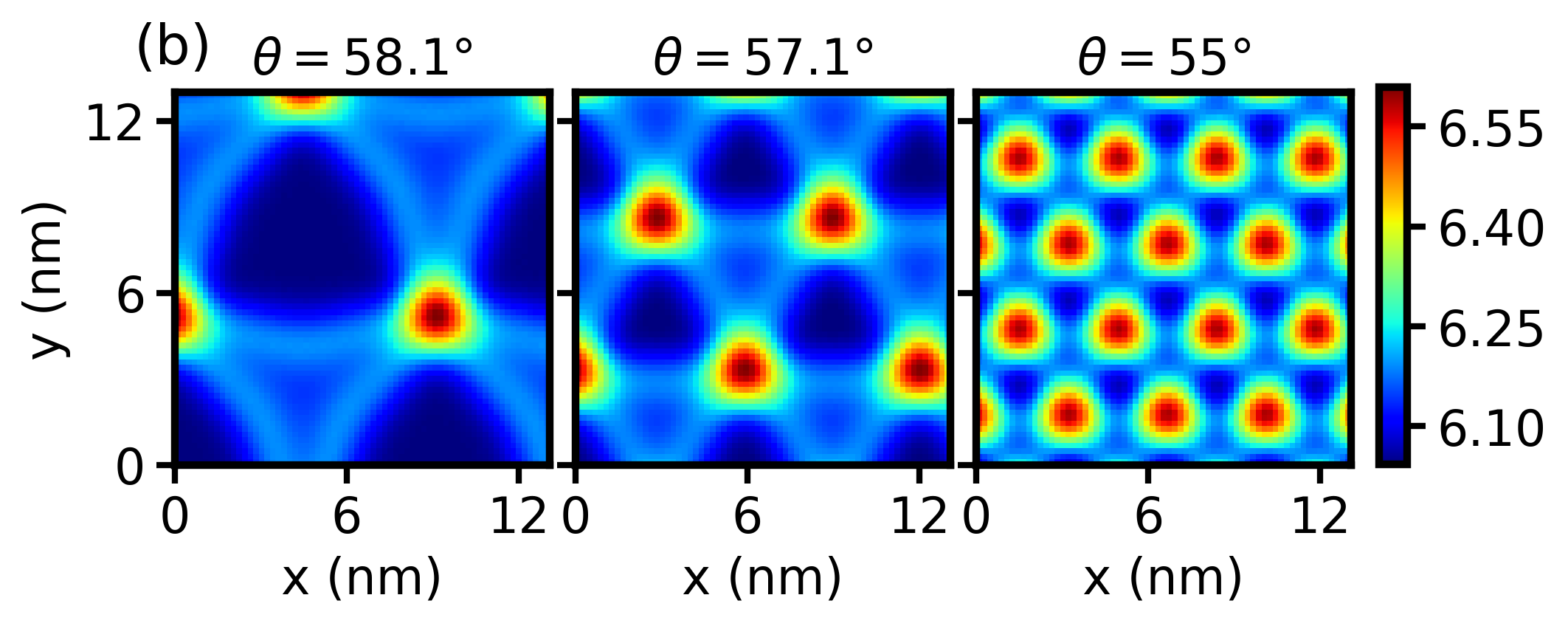}
			\captionsetup{labelformat=empty}
			\caption{}
			\label{fig:rel_ils_b}
		\end{subfigure}

   \vspace{-0.8cm}
   \begin{subfigure}[t]{0.2\textwidth}
   	\hspace{-0.7cm}
   	\includegraphics[width=1.1\textwidth]{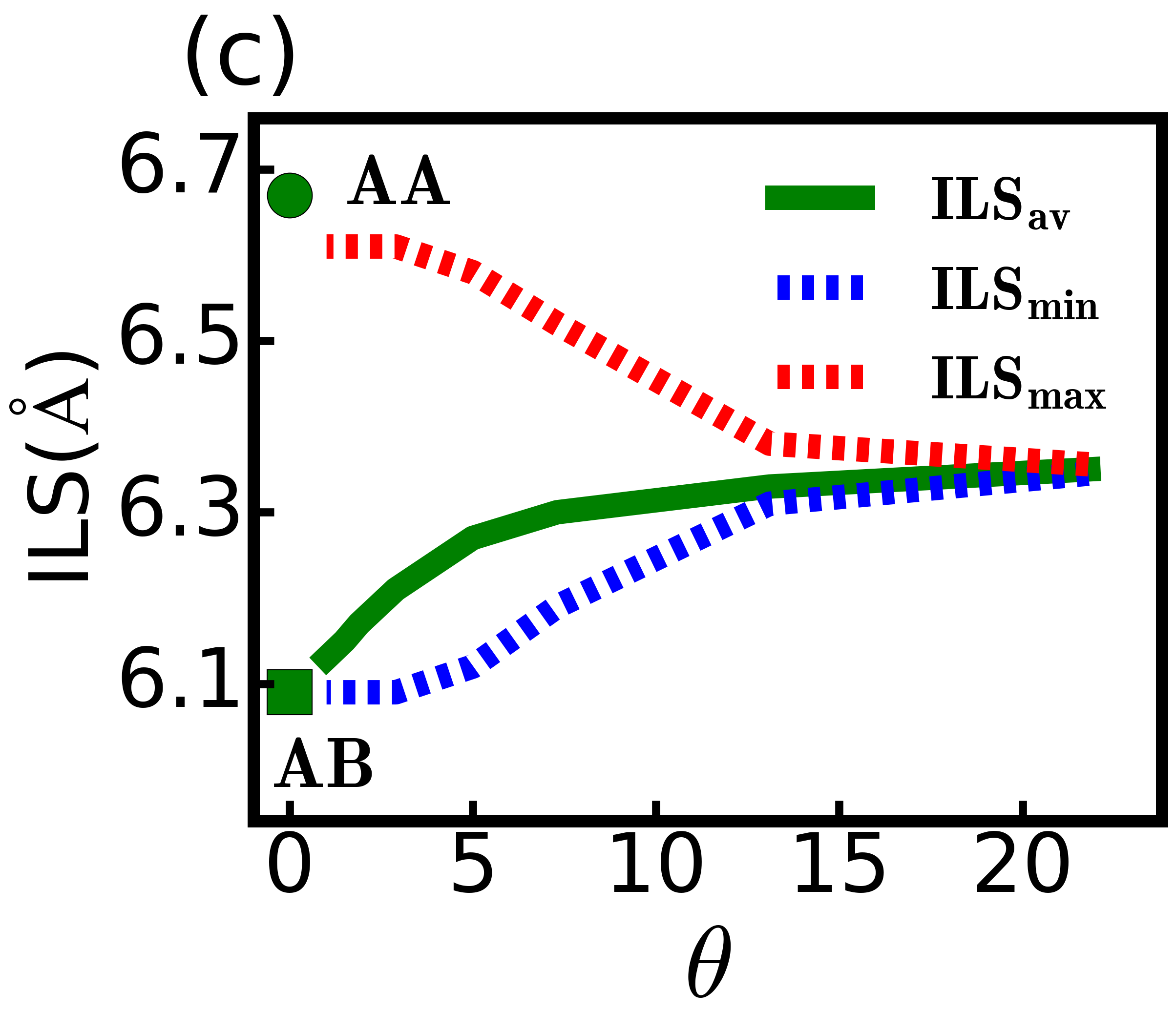}
   	\captionsetup{labelformat=empty}
   	\caption{}
   	\label{fig:rel_ils_c}
   \end{subfigure}
   \hspace{-0.1cm}
   \begin{subfigure}[t]{0.2\textwidth}
   	\includegraphics[width=1.1\textwidth]{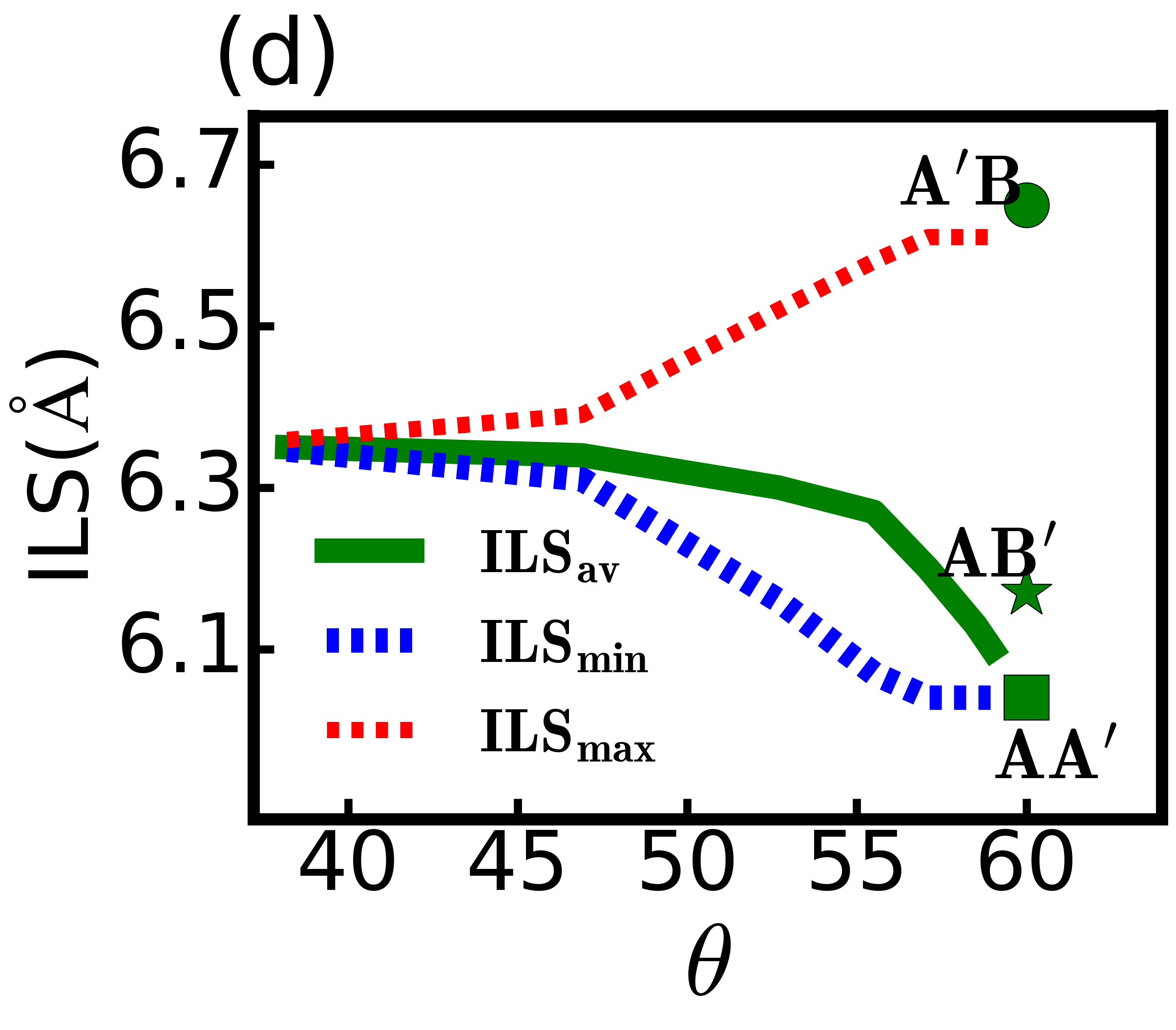}
   	\captionsetup{labelformat=empty}
   	\caption{}
   	\label{fig:rel_ils_d}
   \end{subfigure}
   \vspace*{-0.8cm}
		\caption{Evolution of the ILS landscape and it's average (in $\mathrm{\AA}$) with twist angle in $\mathrm{tBLMoS_{2}}$. The in-plane ($x,y$) distances are in nm.} 
		\label{fig:rel_ils}
	\end{figure}

	\section{\label{sec:level2}Simulation Details}
We use the Twister code \cite{mit_twister} to create the MSL of bilayer $\mathrm{MoS_{2}}$ with several commensurate twist angles $1\degree< \theta<59\degree$. The rigidly twisted structures are relaxed using LAMMPS\cite{Plimpton_jcp_1995, LAMMPS, Bitzek_prl_2006} with the Stillinger-Weber and Kolmogorov-Crespi potentials to capture the intralayer\cite{Jiang_iop_2015, SW} and interlayer interactions of $\mathrm{tBLMoS_{2}}$\cite{mit_kc_2018, KC}, respectively. The interlayer potential consists of two types of interaction: (i) nearest neighbour S-S interaction, and (ii) next nearest neighbour Mo-S interaction\cite{mit_kc_2018}. We use modified PHONOPY \cite{Togo_phonopy, *phonolammps} code to compute the zero temperature vibrational spectra of the relaxed $\mathrm{tBLMoS_{2}}$.
	
Independently, we also compute the low frequency modes from the power spectra of mode projected velocity auto-correlation function (mVACF)\cite{Renata_mVACF,Emmanuel_scirep_2015} from \textit{classical} molecular dynamics simulations with periodic boundary conditions in the canonical ensemble using the Nos\'e-Hoover thermostat in LAMMPS. We equilibrate the system in NVT ensemble for $\sim \ 150-300$ ps at $T=300$ K. The supercell contains 32 MSL in order to correctly capture all the anharmonic effects. Typically, we collect velocities of all atoms in the production run (480 ps, in $\mathrm{NVE}$ ensemble) every 20 timesteps (1 timestep = 1 fs). For computational efficieny, the 480 ps tracjectory is divided into 6 parts, each containing 80 ps. The power spectra of the mVACF projects the full phonon spectra to a particular branch for any $\vec{q}$. The time averaged mVACF at momentum $\vec{q}$ and polarization $s$ is defined as\cite{Renata_mVACF} : 
	 \begin{equation}
	 \langle V_{\vec{q},s}(0) V_{\vec{q},s}^{*}(t) \rangle = \lim_{\tau \to \infty} \frac{1}{\tau} \int_{0}^{\infty} V_{\vec{q},s}(t^\prime) V_{\vec{q},s}^{*}(t+t^\prime) dt^\prime 
	 \end{equation} 
	  with 
	 \begin{equation}
	 V_{\vec{q}, s}(t) = \sum_{j=1}^{N_{t}} \vec{v}_{\vec{q}}^{j}(t).\hat{e}_{\vec{q}, s}^{j}
	 \end{equation}
	  where $j$ denotes the atom type in the unit cell, and $\hat{e}_{\vec{q}, s}$ denotes the eigenvector. The mass weighted momentum projected velocities are defined as, 
	 \begin{equation}
	 \vec{v}_{\vec{q}}^{j}(t) = \sqrt{m_{j}}\sum_{k} e^{-i\vec{q}.\vec{r}_{jk}(t)}\vec{v}_{k}
	 \end{equation}
	 
	  where $\vec{r}_{jk}$ are the atomic coordinates, $k$ denote atoms belonging to particle type $j$ and $m_{j}$ denotes atomic mass. To compute the mVACF, we use the definition of unbiased estimator\cite{Emmanuel_scirep_2015}. Finally, we use fast Fourier transform to compute the power spectra. Instead of computing the mVACF for each eigenmodes in MSL, we use $\mathrm{BLMoS_{2}}$ SM and LBM eigenvectors to compute the mVACF of the superlattice.

	\section{\label{sec:level3}Results}

A $\mathrm{tBLMoS_{2}}$ is composed of different high-symmetry stacking regions, which are different as $\theta \to 0^{\degree}$ and $\theta \to 60^{\degree}$ due to sub-lattice symmtery breaking (Fig.\ref{fig:fig1}). For $\theta \to 0^{\degree}$, there are two unique high-symmetry stacking regions, AA (Mo, S of top layer are directly above Mo, S of bottom layer, respectively) and AB (Bernal stacking with Mo of top layer directly above S of bottom layer, equivalent to BA, also referred to as 3R). For $\theta \to 60^{\degree}$, there are three unique high-symmetry stacking regions, $\mathrm{AA^\prime}$ (Mo, S of top layer are directly above S, Mo of bottom layer, or 2H), $\mathrm{AB^{\prime}}$ (Bernal stacking with Mo of top layer directly above Mo of bottom layer) and $\mathrm{A^\prime B}$ (Bernal stacking with S of top layer directly above S of bottom layer) \cite{Huang_2016_Nanolet}. Among these high-symmetry stackings $\mathrm{AB}$ ($\mathrm{AA^\prime}$) is the most stable with SM frequency $\sim  21\ \mathrm{cm^{-1}}$, whereas $\mathrm{AA}$ ($\mathrm{A^\prime B}$) is unstable with strong imaginary SM frequency. Due to the difference in binding energies of different stackings (consequently, stability and interlayer separation (ILS)), upon relaxing the MSL the more stable stacking regions increase in area. The signatures of the growth of the stable stacking regions with $\theta$ are inherently embedded in the ILS landscape.

\subsection{\label{sec:level3}Relaxation : Interlayer Separation Landscape}
For the calculation of vibrational properties i.e. perturbation with respect to the ground state, the relaxation of the twisted structures are necessary to obtain a suitable ground state. The relaxation of the rigidly twisted structures involves straining and buckling of each constituent layers. In order to demonstrate the consequences of relaxation, in Fig.\ref{fig:rel_ils} we show the $\theta$ dependence of the ILS landscape and $\mathrm{ILS_{av}}$. We can identify the high-symmetry regions in the ILS landscape for $\theta \to 0 \degree$ (Fig.~\ref{fig:rel_ils_a}) : alternate triangles with the least ILS ($\mathrm{AB}$ and $\mathrm{BA}$, deep blue), and red circles with the maximum ILS (AA). Both $\mathrm{AB}$ and $\mathrm{BA}$ regions grow equally for $\theta \to 0^{\circ}$ since they are degenerate in energy. Six domain walls (light blue lines) meet at the ``centers", where AA stacking (topological point defect) regions are located, similar to what happens in twisted bilayer graphene (tBLG) \cite{Alden_PNAS, yoo_natmat_2019, Zhang_2018_jmps, Oleg_2dm_2018}. The ILS landscape for $\theta \to 60 \degree$, on the other hand, (Fig.\ref{fig:rel_ils_b}) is very different due to sub-lattice symmetry breaking. $\mathrm{AA^\prime}$ regions (Reuleaux triangle like) grow overwhelmingly due to their relatively lower binding energy. Six curved domain walls meet at the ``centers", where $\mathrm{A^\prime B}$ (red circles, maximum ILS) stacking regions are located. Generalized stacking fault energy in conjunction with continuum theory is also shown to predict similar in-plane features \cite{Carr_relaxation} for $\mathrm{tBLMoS_{2}}$. It is interesting to note that both the length and the shape of domain walls can be tuned with twists as $\theta\to60\degree$. Figure \ref{fig:rel_ils_c}, \ref{fig:rel_ils_d} capture $\theta$ dependence of $\mathrm{ILS_{av}}$, $\mathrm{ILS_{min}}$ and $\mathrm{ILS_{max}}$ of the MSL. For large twists ($13\degree<\theta<47\degree$), the absence of any extended ideal high-symmetry stacking regions leads to $\theta$-independent behavior of $\mathrm{ILS_{av}, ILS_{min}}$, and $\mathrm{ILS_{max}}$. $\mathrm{ILS_{min}}$ and $\mathrm{ILS_{max}}$ saturate for $\theta < 3\degree$ and $\theta > 57\degree$. Although, $\mathrm{ILS_{av}}$ doesn't saturate in this limit due to the presence of $\mathrm{AA / A^\prime B}$ and domain walls.  
	\par 
	\begin{figure*}[!htbp]
		\vspace{-1cm}
		\begin{subfigure}[t]{0.33\textwidth}
			\captionsetup{labelformat=empty}
			\caption{}
			\hspace{-2cm}
			\label{fig:theta_dep_modes_a}
			\includegraphics[width=0.9\textwidth]{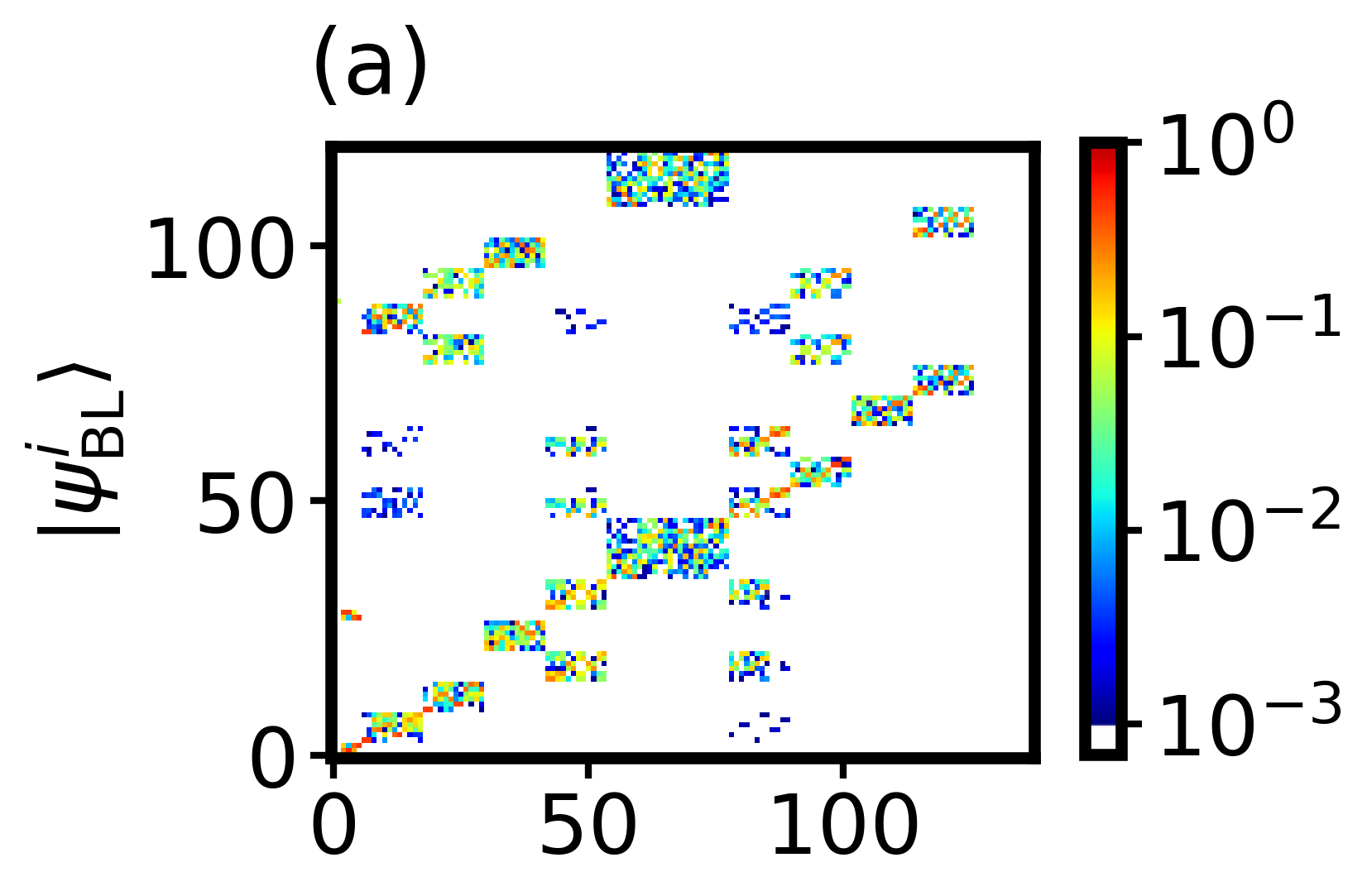}
			\captionsetup{labelformat=empty}
			\caption{}
			\vspace{-1.cm}
			\hspace{-2cm}
			\label{fig:theta_dep_modes_b}
			\includegraphics[width=0.9\textwidth]{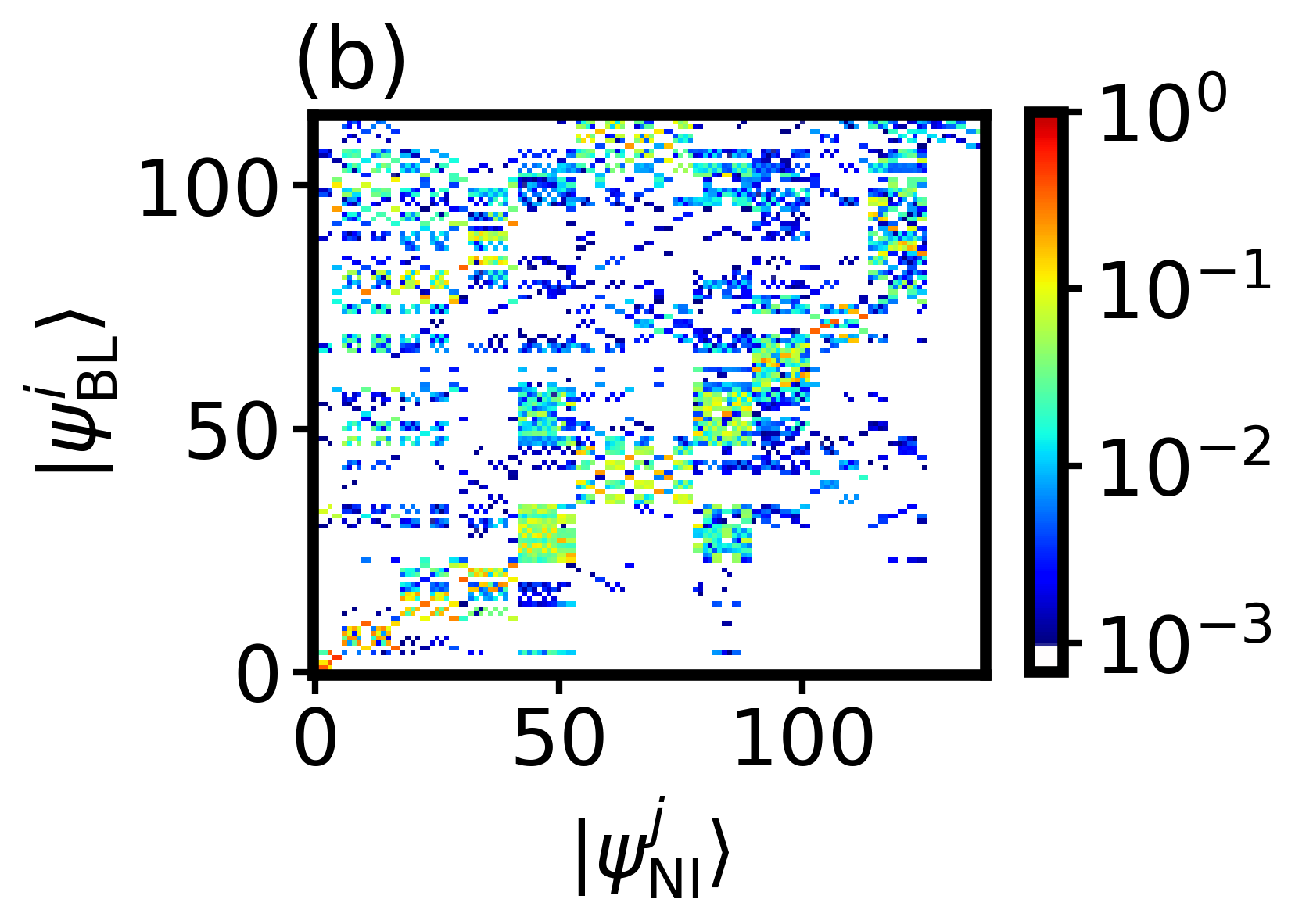}
		\end{subfigure}
		\hspace{-0.7cm}
		\begin{subfigure}[t]{0.33\textwidth}
			\vspace*{0.1cm}
			\captionsetup{labelformat=empty}
			\caption{}
			\label{fig:theta_dep_modes_c}
			\includegraphics[width=1.\textwidth]{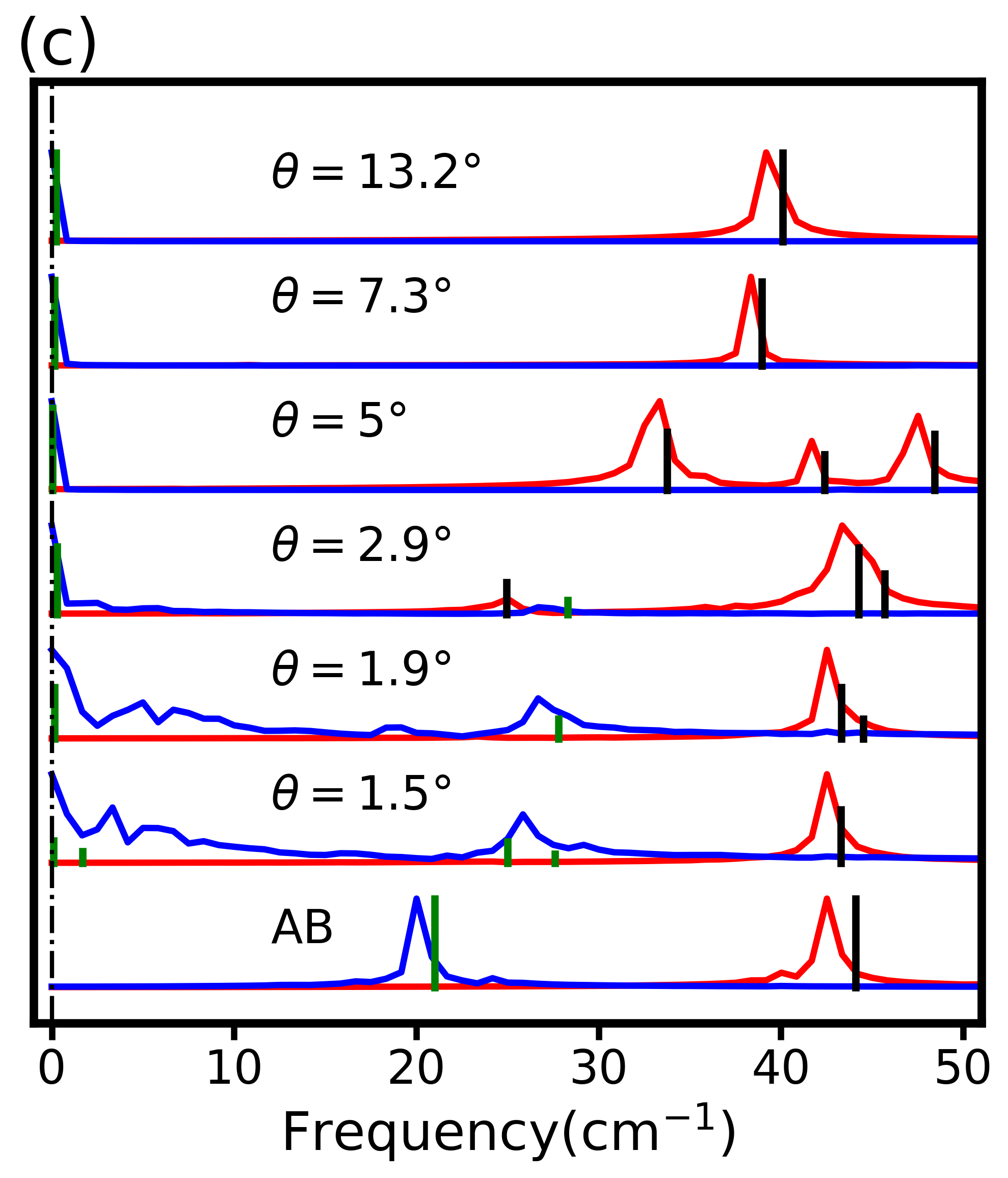}
		\end{subfigure}
		\hspace{-0.1cm}
		\begin{subfigure}[t]{0.33\textwidth}
			\captionsetup{labelformat=empty}
			\caption{}
			\label{fig:theta_dep_modes_d}
			\includegraphics[width=0.98\textwidth]{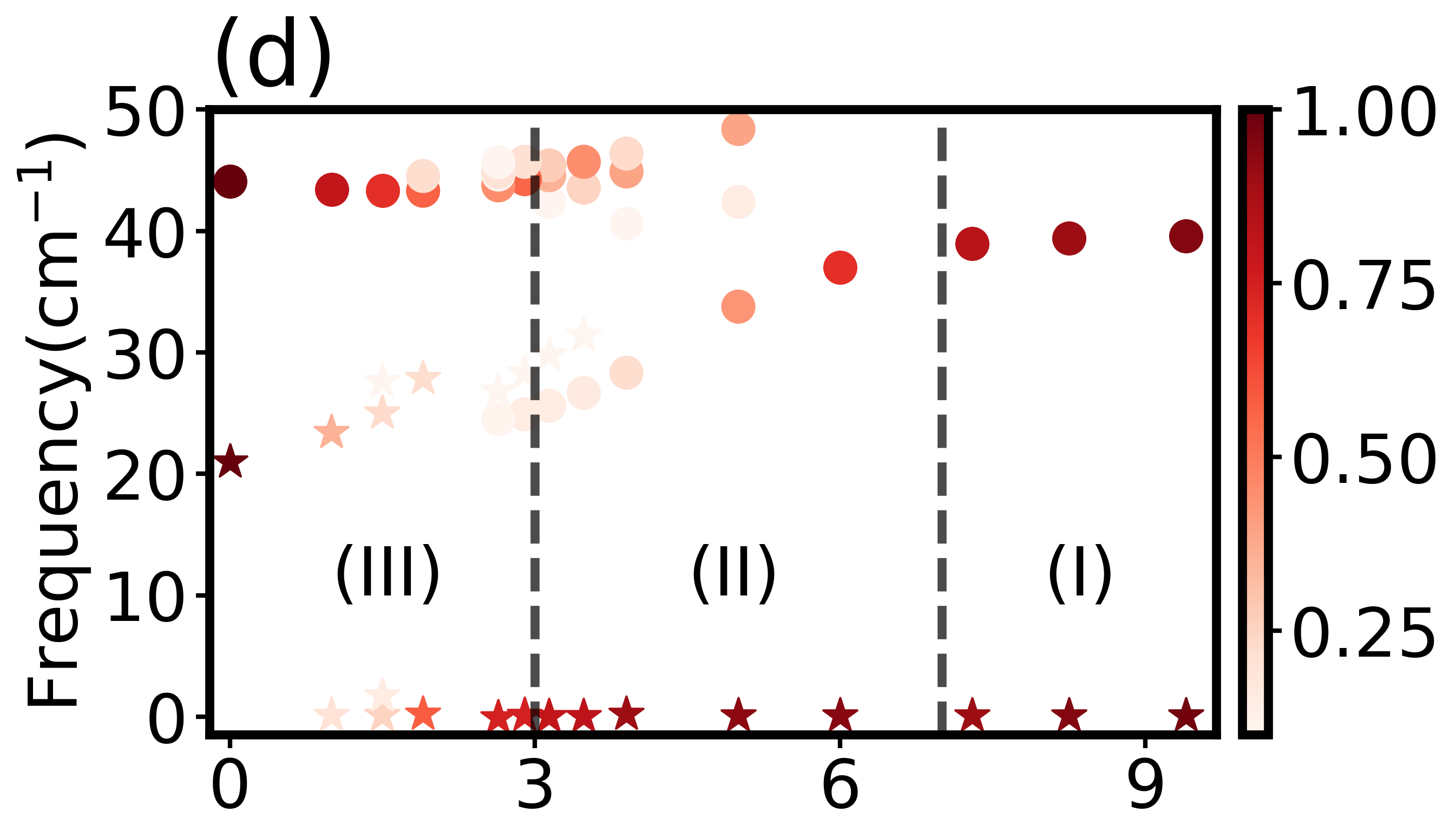}
			\includegraphics[width=0.98\textwidth]{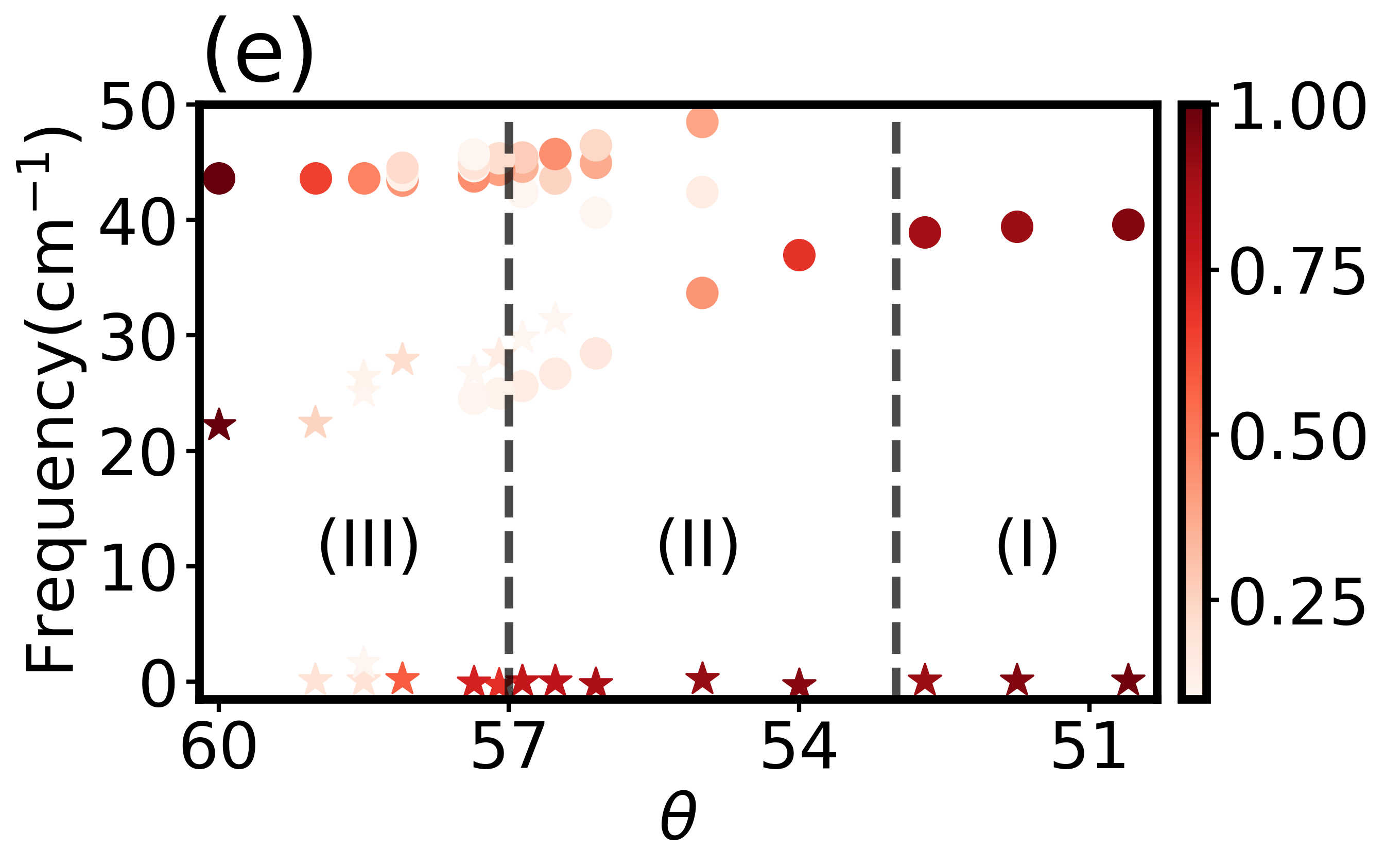}
		\end{subfigure}
		
		\caption{Inhomogeneity in the interlayer coupling in MSL and the evolution of low frequency modes. (a),(b): $A^{ij}$ (see text for definition), for untwisted and twisted ($\theta=57.1\degree$) bilayers of same dimensions, respectively for modes with $\omega < 50 \ \mathrm{cm^{-1}}$. (c): SM (green) (LBM (black)) frequencies at $T=0$ K marked as vertical lines with height proportional to $p^{\mathrm{SM}}$($p^{\mathrm{LBM}}$). Also shown, the power spectra of mVACF (with highest peak normalized to 1) at $T=300$ K. (d),(e): SM and LBM frequencies at $T=0$ K with the colorbar indicating $p^{\mathrm{SM}}, p^{\mathrm{LBM}}$.}
		\label{fig:theta_dep_modes}
	\end{figure*}

\begin{figure}[hbt!]
	\begin{subfigure}[H!]{0.45\textwidth}
		\captionsetup{labelformat=empty}
		\caption{}
		\vspace*{0cm}
		\label{fig:lbm}		
		\includegraphics[width=\textwidth]{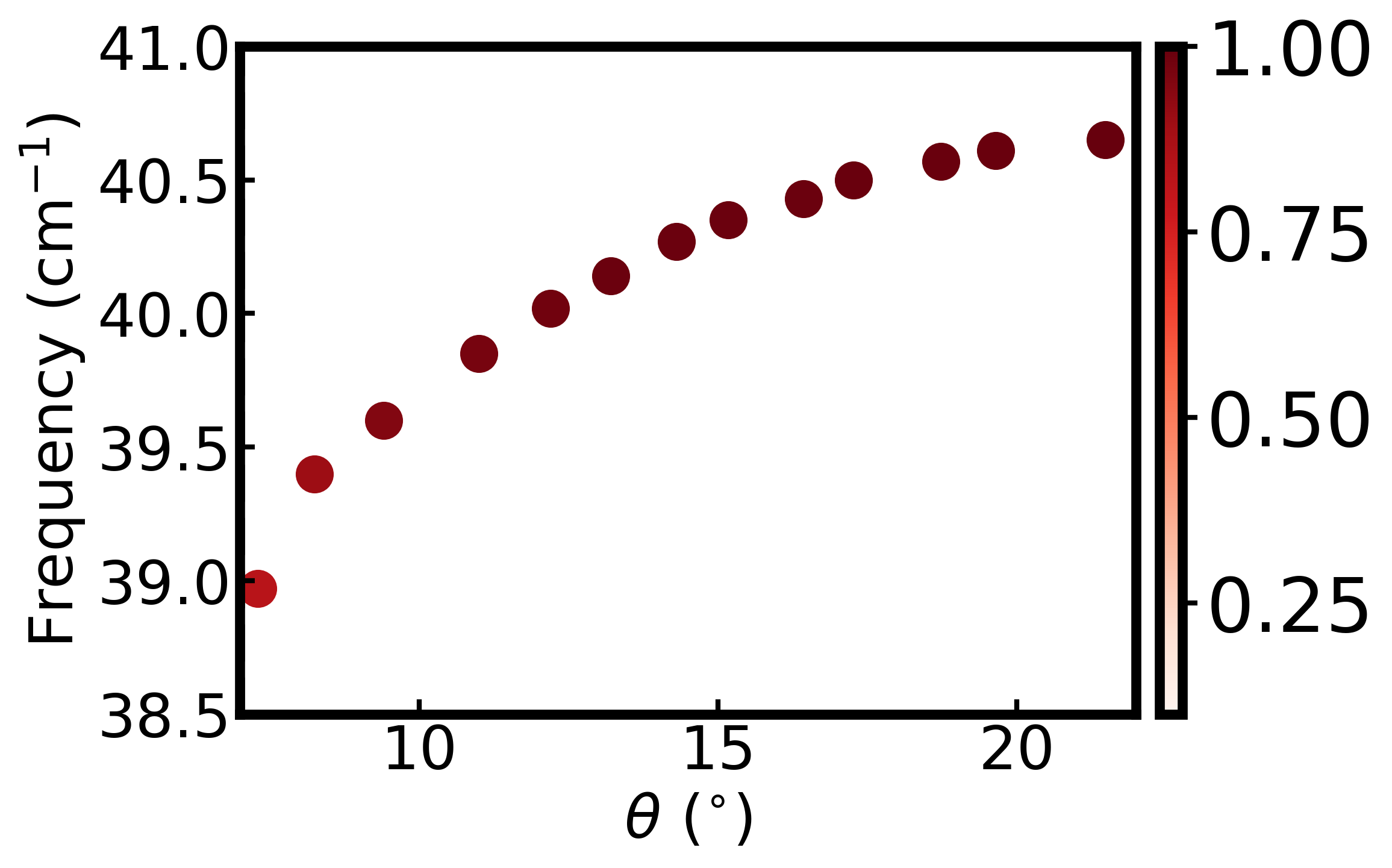}
	\end{subfigure}
	\caption{Monotonic decrement of LBM frequencies of Region (I) as $\theta \to 0\degree$ in $\mathrm{tBLMoS_{2}}$ with the colorbar indicating $p^{\mathrm{LBM}}$. Identical results are obtained near $60^{\circ}$ as well. The calculated LBM frequency for $\mathrm{BLMoS_{2}}$ is 43.5 $\mathrm{cm^{-1}}$. }
	\label{fig:lbm_regI}
\end{figure}

	\begin{figure*}
		\hspace*{-1.5cm}
		\begin{subfigure}[t]{0.95\textwidth}
			\vspace*{-0.4cm}
			\captionsetup{labelformat=empty}
			\caption{}
			\includegraphics[width=\textwidth]{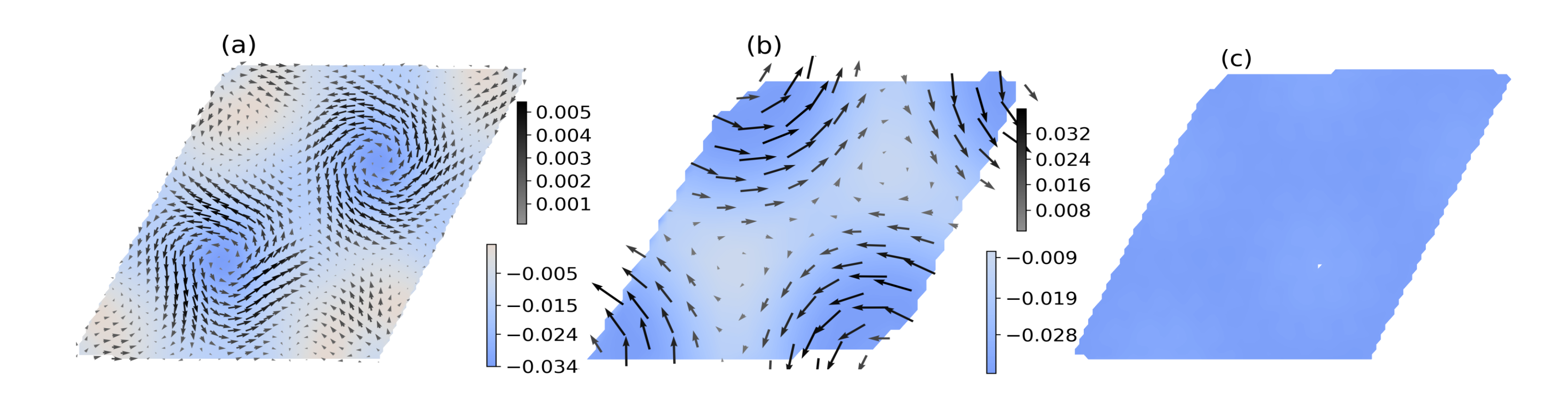}
		\end{subfigure}		
		\vspace*{-0.5cm}
		\caption{\textit{Exact} normalized eigenvectors with the largest projection on bilayer LBM ($p^{\mathrm{LBM}}$) for several $\theta$. The eigenvectors for top Mo layer for $\theta=1.9\degree,\ 5\degree,\ 23.48\degree$ ((a), (b), (c)), respectively; The corresponding bottom layers move exactly in the opposite direction. The arrows indicate the direction and magnitude of the in-plane displacement (with the associated colorbar). The out-of-plane displacement is shown as a field (colored, with associated colorbar). Only at large twist angles, LBM resembles to that of $\mathrm{BLMoS_{2}}$. }.
		\label{fig:lbm_eig}
	\end{figure*}

	\begin{figure*}[!htbp]
		
		\hspace*{-5cm}
		\begin{subfigure}[t]{0.83\textwidth}
			\captionsetup{labelformat=empty}
			\caption{}
			\vspace*{0cm}
			\includegraphics[width=1.35\textwidth, ]{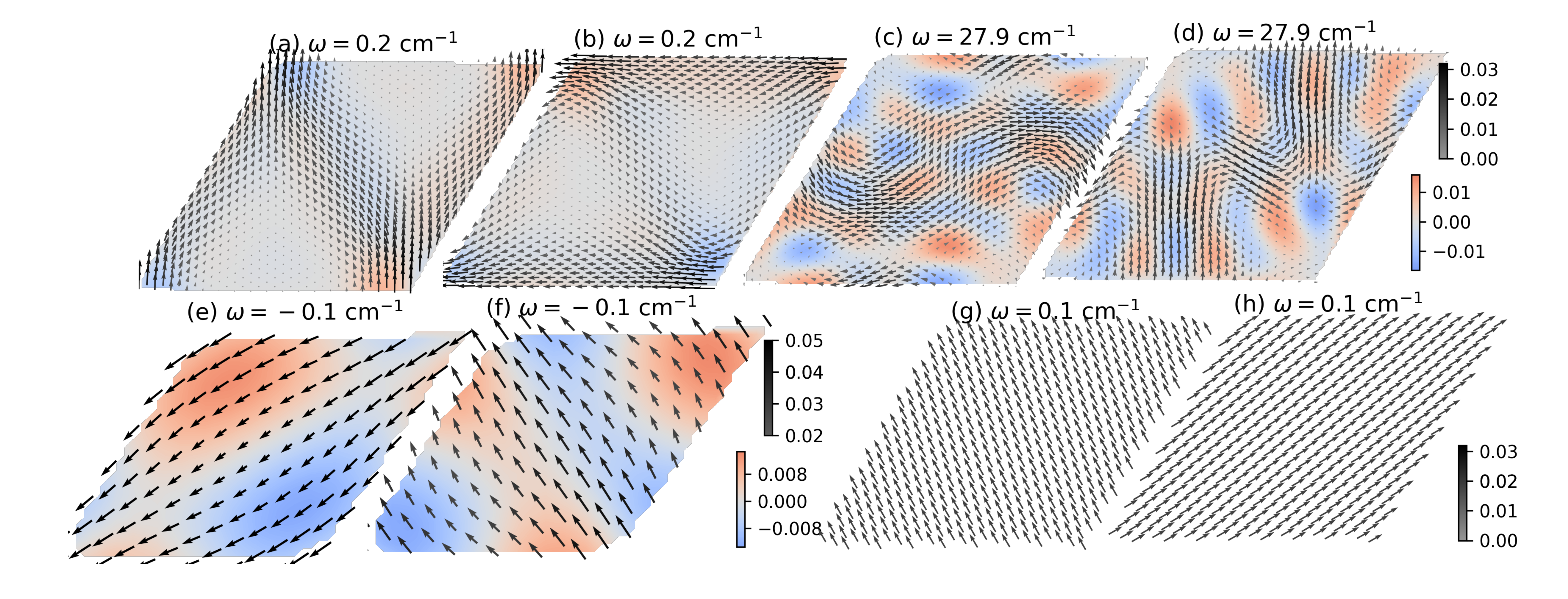}
		\end{subfigure}		
		\caption{A visualization of the eigenvectors at the $\Gamma$ point corresponding to ultra-soft shear modes for $\theta=1.9 \degree$ ((a), (b)), $\theta=5\degree$ ((e), (f)), $\theta=23.48\degree$ ((g),(h)), along with the high frequency SMs ((c), (d)). The position of the domain walls and point defects are same as shown in Fig.~1. The arrows (grey colorbar) denote in-plane displacements (only for Mo atoms of the top layer, for clarity), whereas out-of-plane displacements are represented as a continuous field (colored). The in-plane displacements of the Mo atoms of bottom layer are exactly opposite to that of the top layer.}
		\label{fig:sm_0}
	\end{figure*}

	\begin{figure*}[!htbp]
		
		\hspace*{-3cm}
		\begin{subfigure}[t]{0.9\textwidth}
			\captionsetup{labelformat=empty}
			\caption{}
			\vspace*{0cm}
			\includegraphics[width=1.2\textwidth, ]{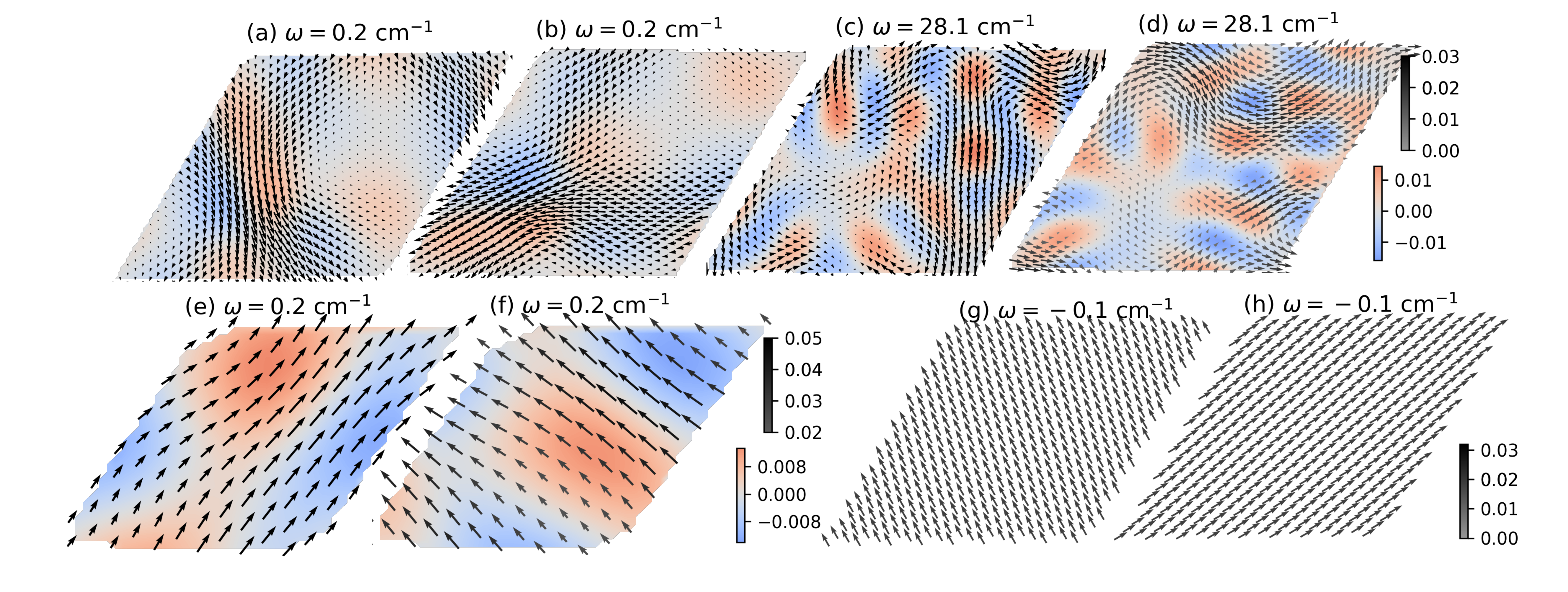}
		\end{subfigure}		
		\caption{A visualization of the eigenvectors at the $\Gamma$ point corresponding to the shear modes for $\theta=58.1 \degree$ ((a), (b)), $\theta=55\degree$ ((e), (f)), $\theta=36.52\degree$ ((g),(h)), along with the high frequency SMs ((c), (d)). The position of the domain walls and point defects are same as shown in Fig.~1. The eigenvectors are denoted in similar manner as in Fig.~6}
		\label{fig:sm_60}
	\end{figure*}

	\subsection{\label{sec:level3}Inhomogeneous Interlayer Coupling}
	As a whole, twisting affects the phonon band structure in two ways. First, the shrinking of the Brillouin zone gives rise to folded phonon modes. Second, as there are multiple stackings in the MSL, the interlayer coupling is inhomogeneous, which leads to mode mixing among in-plane and out-of-plane modes. Relaxation of the rigidly twisted structures further changes this mode mixing and stabilizes the structure (SM frequency of unrelaxed $\mathrm{tBLMoS_{2}}$ is strongly imaginary, implying instability). In order to separate the effects due to only zone folding and the mixing of modes due to presence of multiple stacking, we compute $A^{ij}\ = \ |\langle \psi_{BL}^{i}| \psi_{NI}^{j} \rangle|^{2}$, projections of bilayer eigenmodes ($|\psi_{BL}^{i}\rangle$) onto individual layer modes ($|\psi_{NI}^{j}\rangle$) at $\Gamma$ for both untwisted and twisted structures (Fig.\ref{fig:theta_dep_modes_a}, \ref{fig:theta_dep_modes_b}). The untwisted and twisted structures are of same dimensions. The projections would have been identical if the zone folding effects were the only factor determining phonons in twisted structures. It is clear from the figure that inhomogeneity in the interlayer coupling in the twisted structure, leads to greater mode mixing. The comparison of mode mixing between untwisted and twisted structures are essential as the mixing strongly depends on twist angle and produces non-trivial effects as we illustrate below. This observation is also crucial, as it invalidates the usage of a simple linear chain model to compute SM and LBM frequencies.
	
	\subsection{\label{sec:level3}Shear and Layer Breathing Modes}
	Next, we focus on the effects of twisting on the SM and LBM frequencies with (at $T=300$ K) and without thermal fluctuations ($T=0$ K). While computing the mVACF (for $T=300$ K) we use the SM and LBM eigenvectors of $\mathrm{BLMoS_{2}}$. As discussed above, due to mode mixing the eigenvectors of the twisted structures can be composed of several normal modes of the MSL. Any non-degenerate eigenmode involving relative displacements of the layers of the MSL should appear as a distinct peak in the power spectra of the mVACF. In order to compare with finite $T$ results we also project the eigenmodes of MSL onto the $\mathrm{BLMoS_{2}}$ SM and LBM eigenvectors, $p=|\langle \hat e_{\mathrm{MSL}}|\hat e_{\mathrm{BL}}\rangle|^{2}$ at $T=0$ K (Fig.\ref{fig:theta_dep_modes_c}). It should also be noted that, multiple zone ``folded" modes from SM and LBM branches of different $\vec{q}$ points of the $\mathrm{BLMoS_{2}}$ unit-cell Brillouin zone will appear at the $\Gamma$ point of the MSL Brillouin zone. However, the projections of these ``folded" modes onto $\mathrm{BLMoS_{2}}$ SM and LBM eigenvectors at $\Gamma$ point will be significantly smaller than that of the modes not arising due to zone folding. This is due to orthogonality of the eigenvectors at different $\vec{q}$ points. Therefore, considering the evolution of the eigenmodes with the largest projections (in fact, we consider all modes with $p^\mathrm{{SM}},p^{\mathrm{LBM}}>0.1$ as SM and LBM), we infer the twist angle dependence of shear and layer breathing modes. We can categorize the $\theta$ dependence of the low frequency SM and LBM into three regions. Since the change of the low frequency modes with twist angle is a result of complicated evolution of the interlayer coupling and mode mixing, we present both the frequencies and the eigenvectors of the SM and LBMs. The twist angle dependence of the eigenvectors provide deeper insight to the evolution of the low frequency modes, as we illustrate below. 
	
	  \textbf{Region(I) ($7\degree \lesssim \theta \lesssim 53 \degree$) :} In this region, we find an \textit{averaged} LBM and exceedingly small SM frequencies (0-2 cm$^{-1}$, ``ultra-soft"). The LBM frequency decreases \textit{monotonically} as $\theta \to 7\degree / 53\degree$ from larger twists (Fig.\ref{fig:lbm_regI}). Furthermore, the projections $p^{\mathrm{LBM}}>0.9$, $p^{\mathrm{SM}}>0.9$ indicate that, the nature of vibrations of SM and LBMs remain similar to that of $\mathrm{BLMoS_{2}}$. As a representative of this region, we show the eigenvectors corresponding to SM and LBM in Fig.{\ref{fig:sm_0}g,\ref{fig:sm_0}h,\ref{fig:sm_60}g,\ref{fig:sm_60}h,\ref{fig:lbm_eig}c} for $\theta=23.48^{\circ}$. Clearly, the SM (LBM) eigenvectors correspond to the relative horizontal (vertical) uniform displacement of the constituent layers. This is due to the absence of any extended high-symmetry stacking in the MSL, which leads to nearly uniform interlayer coupling. However, as we approach Region(II), the LBM starts mixing with in-plane modes giving rise to the \textit{monotonic} decrease in the LBM frequencies. The change in LBM frequencies (by $\sim 1.5\ \mathrm{cm^{-1}}$) can be used to reliably infer large $\theta$. Moreover, the presence of LBM also indicates that the layers in the twisted structures are not completely decoupled (in the out-of-plane direction). 
	  
	  \textbf{Region(II) ($3\degree \lesssim \theta \lesssim 7\degree$, $53\degree \lesssim \theta \lesssim 57\degree$) :} We find the LBM frequencies are quite sensitive to twist angle including the presence of multiple LBMs. The eigenvectors corresponding to LBMs clearly indicate the mixing with in-plane modes. As an example, we show the LBM eigenvector with maximum $p^{\mathrm{LBM}}$ for $\theta=5^{\circ}$ (Fig.{\ref{fig:lbm_eig}b}). The in-plane modes corresponding to the LBM is of similar order in magnitude. In this region all the high-symmetry stackings, and domain walls occupy comparable area-fraction of the MSL. This enhances the mode mixing. The ultra-soft SMs are also present in this region. However, the corresponding SM eigenvectors are no longer completely uniform (Fig.\ref{fig:sm_0}e,\ref{fig:sm_0}f,\ref{fig:sm_60}e,\ref{fig:sm_60}f) and start mixing with out-of-plane modes. In short, this region represents the transition from completely mismatched stacking (Region I) to highly ordered stable stacking regions separated by domain walls (Region III).

	\textbf{Region(III) ($\theta\lesssim 3\degree,\theta \gtrsim57\degree$) :} As $\theta$ decreases further ($\theta \to 0^{\circ}$, or $\theta \to 60^{\circ}$) we find one LBM (with significantly large $p^{\mathrm{LBM}}$) with frequency similar to that of stable $\mathrm{BLMoS_{2}}$. This is due to the overwhelming growth of the stable stacking regions (Fig.\ref{fig:rel_ils}) in this region. The mixing of LBM with in-plane modes although exists, are smaller compared to Reg.II (Fig.\ref{fig:lbm_eig}). In order to establish this, we plot the LBM eigenvector with largest $p^{\mathrm{LBM}}$ for $\theta=1.9^{\circ}$ (Fig.\ref{fig:lbm_eig}a). It is evident from the figure that, the LBM primarily arises from the out-of-plane vibration of the $\mathrm{AB}$ stacked region. On the other hand, the SM frequencies split essentially in two branches : one ultra-soft in nature (similar to Region I, II) and one with high frequency (similar to $\mathrm{BLMoS_{2}}$, 22-28 $\mathrm{cm^{-1}}$). Moreover, as $\theta$ decreases from $3^{\circ}$ to $0^{\circ}$ ($57^{\circ}$ to $60^{\circ}$) the high-frequency SM redshifts (Fig.\ref{fig:theta_dep_modes_d},3e). The splitting of the SM has important consequences on the frictional properties, as we discuss later. The ultra-soft SMs are localized on domain walls and $\mathrm{AA}$ stacking (domain walls and $\mathrm{A^\prime B}$ near $60^{\circ}$, Fig.\ref{fig:sm_0}a, \ref{fig:sm_0}b, \ref{fig:sm_60}a, \ref{fig:sm_60}b). On the contrary, high-frequency SMs primarily originate from the relative displacement of stable stacking regions (Fig.\ref{fig:sm_0}c, \ref{fig:sm_0}d, \ref{fig:sm_60}c, \ref{fig:sm_60}d). The splitting of SM frequencies occurs due to significant growth of the stable stacking region in the $\mathrm{tBLMoS_{2}}$. The high-frequency SM are quite similar in magnitude as $\mathrm{AB}$ ($\mathrm{AA^\prime}$). The apparent stiffening of these modes as $\theta  \to 3^{\circ}$ is because of mixing with out-of-plane modes. For instance, the SM frequency of individual $\mathrm{AB}$ stacking is $\sim 21 \ \mathrm{cm^{-1}}$, whereas the high-frequency SM of $\mathrm{tBLMoS_{2}}$ with $\theta = 3^{\circ}$ is $\sim 28 \ \mathrm{cm^{-1}}$.  It must be pointed out again, that the evolution of SMs does not arise from the folded modes of the unit-cell Brillouin zone as discussed earlier. The twist angle dependence of these eigenvectors are strongly dependent on the interlayer interaction strength. To further illustrate this point, we compare the ultra-soft SM eigenvectors for $\mathrm{tBLMoS_{2}}$ with $\theta=23.03^{\circ}$ (MSL lattice constant $\sim 47.3$ {\AA} with 5418 atoms) and $\theta=1.9^{\circ}$ (MSL lattice constant $\sim 47.7$ {\AA} with 5514 atoms). The ultra-soft SM eigenvectors in these two systems are remarkably different although they are of similar dimensions. For $\theta=23.03^{\circ}$ the eigenvectors represent uniform relative displacement of the layers, whereas for $\theta=1.9^{\circ}$ the eigenvectors are localized. This shows that, the eigenvectors corresponding to the ultra-soft SMs are controlled by the interlayer interaction strength and can not be explained by simple zone folding argument. Moreover, the high-frequency SMs are only present for $\theta=1.9^{\circ}$ and absent for $\theta=23.03^{\circ}$, further confirming our conclusion. The maximum variation in the SM and LBM frequencies ($\sim 8$ $\mathrm{cm^{-1}}$ and $\sim 10$ $\mathrm{cm^{-1}}$) are comparable to those observed in Raman studies ($\sim 8 \ \mathrm{and} \ 6.7 \ \mathrm{cm^{-1}}$, respectively \cite{Huang_2016_Nanolet}). Such large variations of low frequency modes, appearance of multiple LBMs can be useful for characterization of bilayer properties.
	
	\par

	\begin{figure}
		\hspace*{-0.5cm}
		\begin{subfigure}[t]{0.55\textwidth}
			\vspace*{-1.4cm}
			\captionsetup{labelformat=empty}
			\caption{}
			\includegraphics[width=\textwidth, scale=0.5]{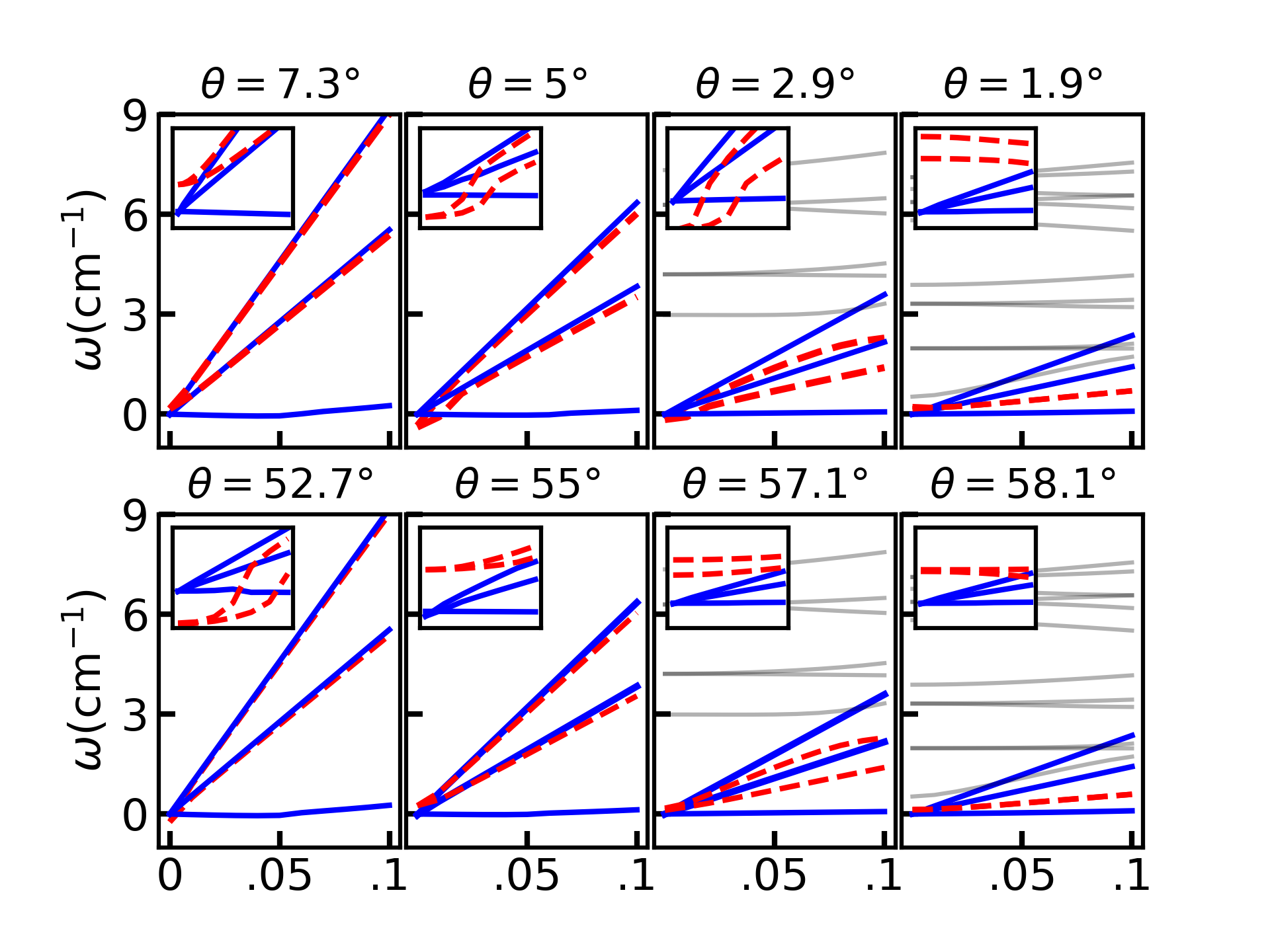}
		\end{subfigure}		
		\vspace*{-0.5cm}
		\caption{Dispersion of the low frequency modes for several $\theta$. The $x$ axis represents momentum, in units of $\frac{4\pi}{\sqrt{3} \ a_\mathrm{m}}$, with $a_{\mathrm{m}}$ being the moir\'{e} lattice constant. The LA, TA and ZA (ultra-soft shear) modes are highlighted with blue (red-dashed) lines. Insets show zoomed in dispersion for $q<0.005$ with $\omega < 0.4 \ \mathrm{cm^{-1}}$.}.
		\label{fig:disp_sm}
	\end{figure}
	
	\begin{figure}
		\hspace*{-0.6cm}
		\begin{subfigure}[t]{0.25\textwidth}
			\vspace*{-1cm}
			\captionsetup{labelformat=empty}
			\caption{}
			\includegraphics[width=\textwidth]{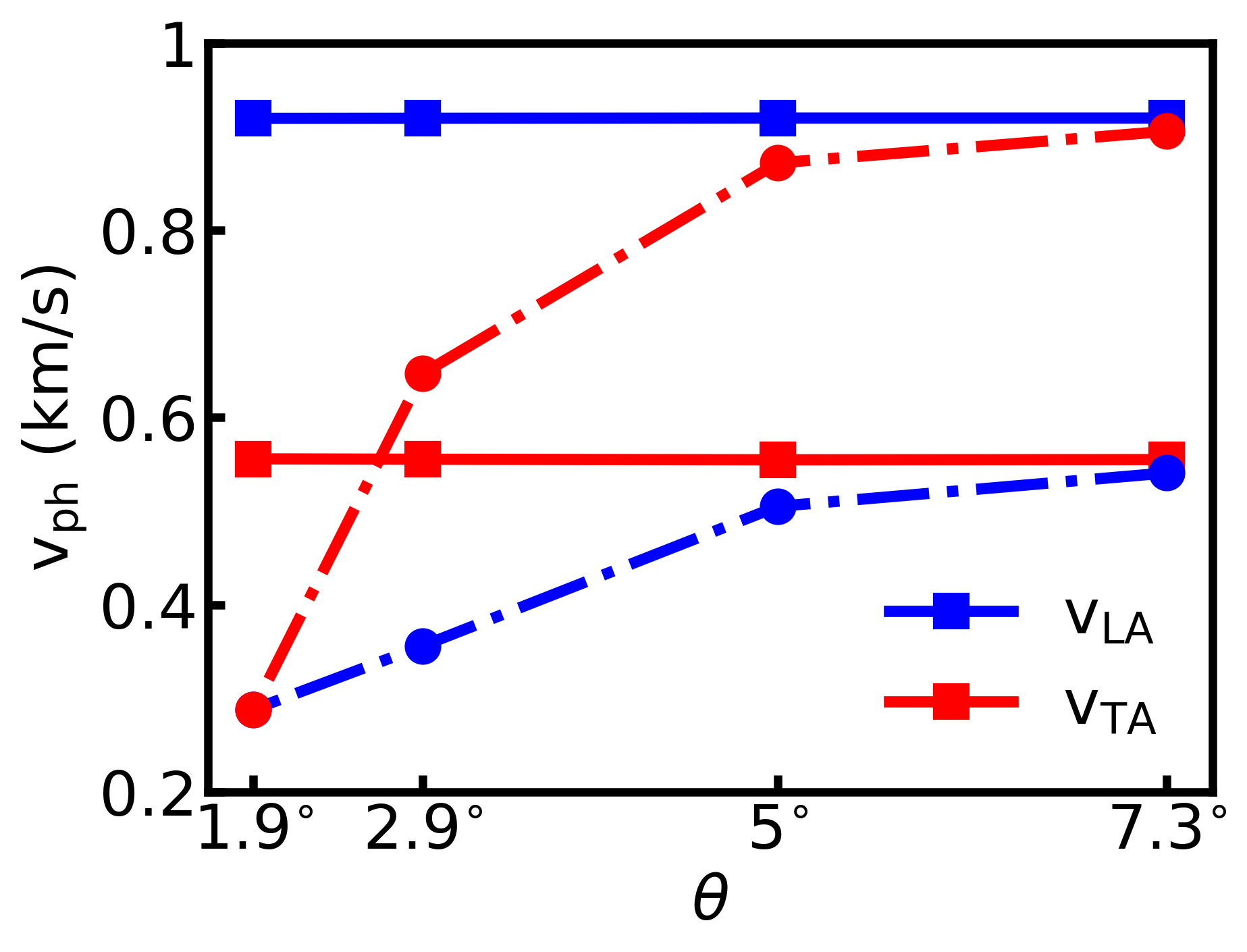}
		\end{subfigure}	
		\begin{subfigure}[t]{0.25\textwidth}
			\vspace*{-1cm}
			\captionsetup{labelformat=empty}
			\caption{}
			\includegraphics[width=\textwidth]{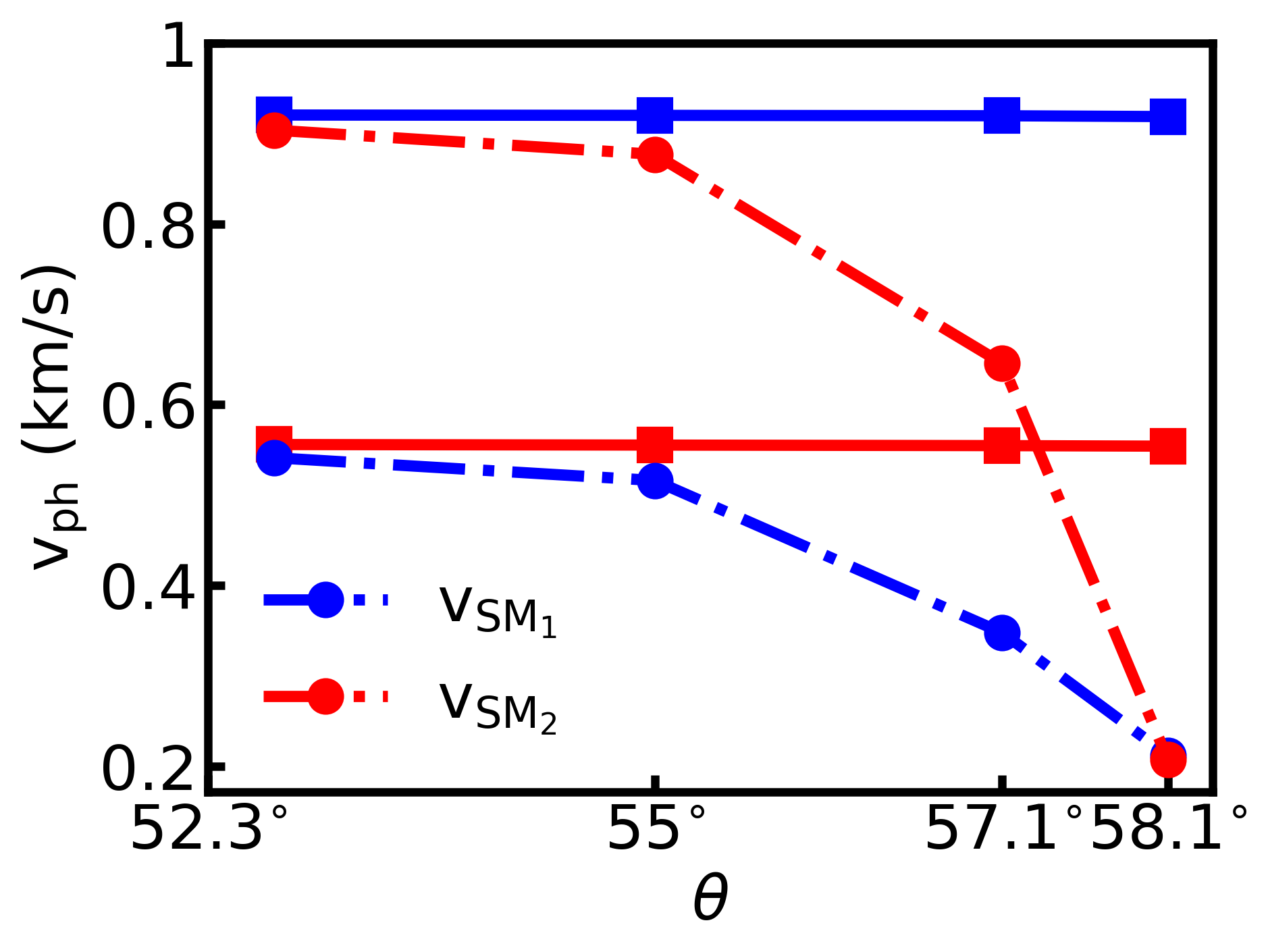}
		\end{subfigure}				
		\vspace*{-0.1cm}
		\caption{Twist angle dependence of the velocity of the phonon modes for both acoustic modes (LA, TA) and ultra-soft phason modes computed from the linear region of phonon dispersion.}.
		\label{fig:vel_ph}
	\end{figure}	
\subsection{\label{sec:level3}Ultra-soft Shear Modes : Phasons}	
The presence of the ultra-soft modes with twist angle is one of the major finding of our work. Thus, we investigate three important aspects of this finding : the origin of these modes, their twist angle dependence and consequences on frictional properties. 

The ultra-soft modes represent an \textit{effective} translation of the MSL by local relative displacements of the atoms in the constituent layers (see attached movies). Since the frequencies associated with these modes are also very small (almost acoustic mode like), this implies that under the relative displacements of two layers (following the eigenvectors shown in Fig.\ref{fig:sm_0},\ref{fig:sm_60}) the energy of the $\mathrm{tBLMoS_{2}}$ remains invariant (or almost invariant), which can be a consequence of continuous symmetry breaking. For example, the in-plane acoustic modes (LA, TA modes) which represent invariance of the total energy under global translation originate due to translational symmetry breaking. However, strictly speaking, we show that the ultra-soft SMs found in our calculation are optical modes and does not represent continuous symmetry breaking. The optical nature is clearly reflected in the dispersion relation ($d\omega / dq \approx0$, for small $q$ very near to $\Gamma$ point) unlike the acoustic modes (Fig.\ref{fig:disp_sm}, small negative values at $\Gamma$ , $\lesssim -0.2 \ \mathrm{cm^{-1}}$, are within numerical accuracies of our calculation). We highlight both the acoustic and ultra-soft modes in order to show this difference clearly. The ultra-soft nature of these SMs can be understood from one-dimensional Frenkel-Kontorova model. In this model, a linear chain of atoms is subjected to an external periodic potential. Depending on the ratio of the periodicity of the potential and linear chain, the structures can be commensurate or incommensurate. For simplicity, we assume the incommensurate (commensurate) structure as a lattice with infinite (finite, small) periodic length. Without considering dissipative coupling, the incommensurate structure possesses a gapless Goldstone mode (\textit{phason}) with linear dispersion due to invariance of the phases of two mass density waves under uniform relative displacement. When a commensurate state is approached, the phason becomes gapped\cite{Chaikin_book_1995} with $d\omega / dq \approx0$. This is exactly what happens in the case of ultra-soft SMs found in our calculations. The linear chain of atoms and the external periodic potential in the Frenkel-Kontorova model are replaced by the constituent $\mathrm{MoS_{2}}$ layers of $\mathrm{tBLMoS_{2}}$, and stacking dependent binding energy, respectively. Since we only simulate commensurate angles, the \textit{phasons} are always gapped (although ultra-soft) optical modes with $d\omega / dq \approx0$. Hence, in our calculation the phase invariance associated with these modes are always approximate. In the case of incommensurate twist angles, however, the phase invariance becomes exact and we expect corresponding gapless phason modes with $\omega \propto q$.

The exact eigenvectors for the ultra-soft SMs can not be directly probed by experiments. Keeping in mind the possible experimental signatures of twist angle dependence, we compute the $d\omega / dq$ (group velocity) of these ultra-soft modes. We compare them with the velocities of the acoustic modes (Fig.\ref{fig:vel_ph}) away from $\Gamma$ point using the linear dispersion of these modes. The velocity of the acoustic modes remains invariant with respect to the change of twist angle, whereas the velocity of the ultra-soft phason modes change significantly (by a factor of 2-3 at $\theta=1.9^{\circ},58.1^{\circ}$). Physically, the in-plane acoustic modes represent the long-wavelength vibrations of the entire twisted bilayer $\mathrm{MoS_{2}}$, with all the atoms of two layers participating. At $\Gamma$ point, these modes correspond to uniform translation of all the atoms in the MSL. Therefore, the acoustic mode velocities within the harmonic approximation can be written in terms of the Lam\'{e} coefficients, $\mu, \lambda$ in the following manner : $\mathrm{v_{LA}}=\sqrt{(\lambda + 2\mu) / \rho}$ and $\mathrm{v_{TA}}=\sqrt{\mu / \rho}$ ~\cite{Chaikin_book_1995}. The presence of the MSL only affects the motion involving relative displacement of the constituent layers. Since, the LA and TA modes only correspond to ``in-phase" motion of the layers, they remain unaffected irrespective of twist angle. Moreover, the rigidity (governed by $\lambda, \ \mu$) of the lattice for the in-plane acoustic modes are the same as that of single layer $\mathrm{MoS_{2}}$. This is evident from the twist angle independent behavior of the velocities of the LA, TA modes (Fig.\ref{fig:vel_ph}). On the other hand, the ultra-soft phason modes are localized on domain walls and $\mathrm{AA}$ stacking for $\theta \to 0^{\circ}$ (domain walls and $\mathrm{A^\prime B}$ stacking). They represent ``acoustic" modes of the emergent soft moir{\`e} scale lattice (contains only domain walls and $\mathrm{AA}$ near $0^{\circ}$, whereas domain walls and $\mathrm{A^\prime B}$ near $60^{\circ}$ ). Strictly speaking, they become acoustic modes only for incommensurate structures, as pointed out earlier. The soft nature of the lattice is noticeable from the change of velocity of these modes (Fig.\ref{fig:vel_ph}). Similar to the LA, TA modes the velocities of the phason modes are functions of $\lambda_{E}, \ \mu_{E}$, where $\lambda_{E}, \ \mu_{E}$ represents effective Lam\'{e} coefficients of the moir{\`e} scale lattice. The decrease in ultra-soft modes velocity imply reduction in strength of the effective Lam\'{e} coefficients. This is because ultra-soft modes originate from the ``out-of-phase" motion of the two layers. Thus, the strength of the interlayer coupling gives rise to effective Lam\'{e} coefficients.  We leave details of the consequences of this emergent soft lattice and investigation of their rigidity for a subsequent work. 



	\begin{figure*}
		\hspace*{-0.5cm}
		\begin{subfigure}[t]{1\textwidth}
			\vspace*{-0.3cm}
			\captionsetup{labelformat=empty}
			\caption{}
			\includegraphics[width=\textwidth]{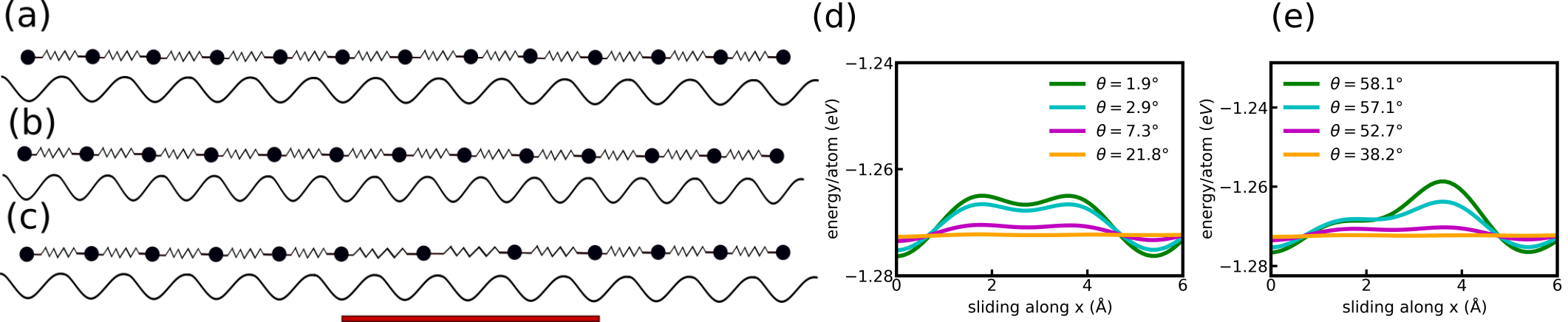}
		\end{subfigure}		
		\vspace*{-0.1cm}
		\caption{(a),(b),(c): One dimensional Frenkel-Kontorova model with unit-cell commensuration, large-scale commensuration, mostly unit-cell commensurate with discommensuration in between. The discommensuration is marked with a red line. (d),(e): Development of barrier against shearing as $\theta \to 0\degree/60\degree$. After relaxation of the twisted structures we slide the top $\mathrm{MoS_{2}}$ layer along the X axis and calculate the total energy without further relaxation.}.
		\label{fig:barrier}
	\end{figure*}

 Before discussing the consequence of the twist angle dependence of the ultra-soft SM frequencies on frictional properties, we briefly summarize previous experimental studies on this aspect. The introduction of a twist between two layers of two dimensional materials has been shown to modify the frictional properties dramatically\cite{Ribeiro_rotatable_science_2018,Koren_prb_2016,Oded_prb_2012,Dien_2004_PRL,Li_am_2017,Wang_prb_2019,Koren_scaling_prb_2016,Leven_jpcl_2012,Rasool_nanolett_2019,Victor_cms_2019}. Depending on the twist angle, the frictional force can be exceedingly small (\textit{structural superlubricity}) or moderately large. For instance, the frictional force can drop by 2-3 orders of magnitude if the twist angle is $>5^{\circ}$ or $<55^{\circ}$ in the case of graphene on graphite, twisted bilayer $\mathrm{MoS_{2}}$\cite{Dien_2004_PRL,Li_am_2017,Onodera_jpcb_2010,Ribeiro_rotatable_science_2018}. Also, for $5^{\circ} \lesssim \theta \lesssim55^{\circ}$, the frictional properties remain almost constant. In all these examples, maximum friction is obtained when the two layers are untwisted. Although microscopic details of the interlayer sliding process giving rise to superlubric behavior can be quite involved, the primary reason behind superlubricity is often attributed to incommensurability between two surfaces. Here, we identify that the change of frictional properties with $\theta$ are intimately related to the existence and evolution of the ultra-soft modes. For $\theta=0^{\circ} / \mathrm{60^{\circ}}$, the $\mathrm{BLMoS_{2}}$ is unit cell commensurate. When trying to shear one layer with respect to another in $\mathrm{BLMoS_{2}}$, all unit cells have to cross the interlayer-sliding barrier simultaneously. This leads to large SM frequencies ($\sim 21 \ \mathrm{cm^{-1}}$) and high-friction. This can also be understood from the aforementioned one dimensional Frenkel-Kontorova model. In the unit-cell commensurate case (Fig.{\ref{fig:barrier}a}), while shearing the atoms globally with respect to the external periodic potential, every atom rises toward the hill. Thus, the cost of shearing is large. However, if the system is large-scale periodic (Fig.{\ref{fig:barrier}b}), then every atom has to cross variable sliding barrier. This reduces the shearing energy, due to cancellation of cost while rising up the hill and gain while falling from it. Similarly, $\mathrm{tBLMoS_{2}}$ for large twist angles ($\theta \lesssim 5^{\circ}, \gtrsim 55^{\circ}$) is periodic at larger scale. While shearing unit cells have to cross a variable interlayer-sliding barrier (due to coexistence of multiple stackings with different binding energies). In effect, this drastically reduces SM frequencies and hence friction as well (superlubric). This can be easily confirmed from the existence of the ultra-soft phason modes and the absence of high-frequency SM. However, as $\theta$ decreases further not only the high-frequency SM similar to the untwisted bilayer appear, but also the ultra-soft modes start to localize on the domain walls (splitting of SM). This immediately implies that, to shear one layer with respect to the other globally a large barrier has to be overcome. Because, only a small fraction of atoms participate in the ultra-soft modes in this case unlike large twist angle. A significant fraction of atoms participate in the high-frequency SM implying pinning. The origin of the pinning lies in the significant growth of stable stacking when $\theta \to 0\degree/60\degree$ (Fig.\ref{fig:rel_ils}) leading to the development of large interlayer sliding barrier against shearing (Fig.\ref{fig:barrier}d,\ref{fig:barrier}e). Similarly, if the potential is strong enough in the Frenkel-Kontorova model, most atoms sit at the minima. Only a small number of atoms occupy the energetically unfavorable hills, known as discommensuration (Fig.\ref{fig:barrier}c). This leads to pinned phasons in the Frenkel-Kontorova incommensurate structures. The nature of vibrations of ultra-soft phason modes in our calculations (uniform at large $\theta$ due to complete stacking mismatch and non-unifrom at small $\theta$ due to overwhelming growth of stable stacking) indicate the possibility of having \textit{pinned phasons} (Aubry-like transition) in the small twist incommensurate structures \cite{Chaikin_book_1995, Lubensky_prb_1985, Peyrard_iop_1983,Bak_rpp_1982}. Interestingly, similar pinning behavior has also been realized in systems like colloidal monolayers in optical lattices \cite{Brazda_prx_2018, Mandelli_prb_2015}, and physisorbed sub-monolayers on crystal surfaces\cite{Pierno_natnano_2015}. 

\section{\label{sec:level4}Discussions}

\begin{figure}
		\hspace*{-1cm}
		\begin{subfigure}[t]{0.44\textwidth}
			\vspace*{-0.3cm}
			\captionsetup{labelformat=empty}
			\caption{}
			\includegraphics[width=\textwidth]{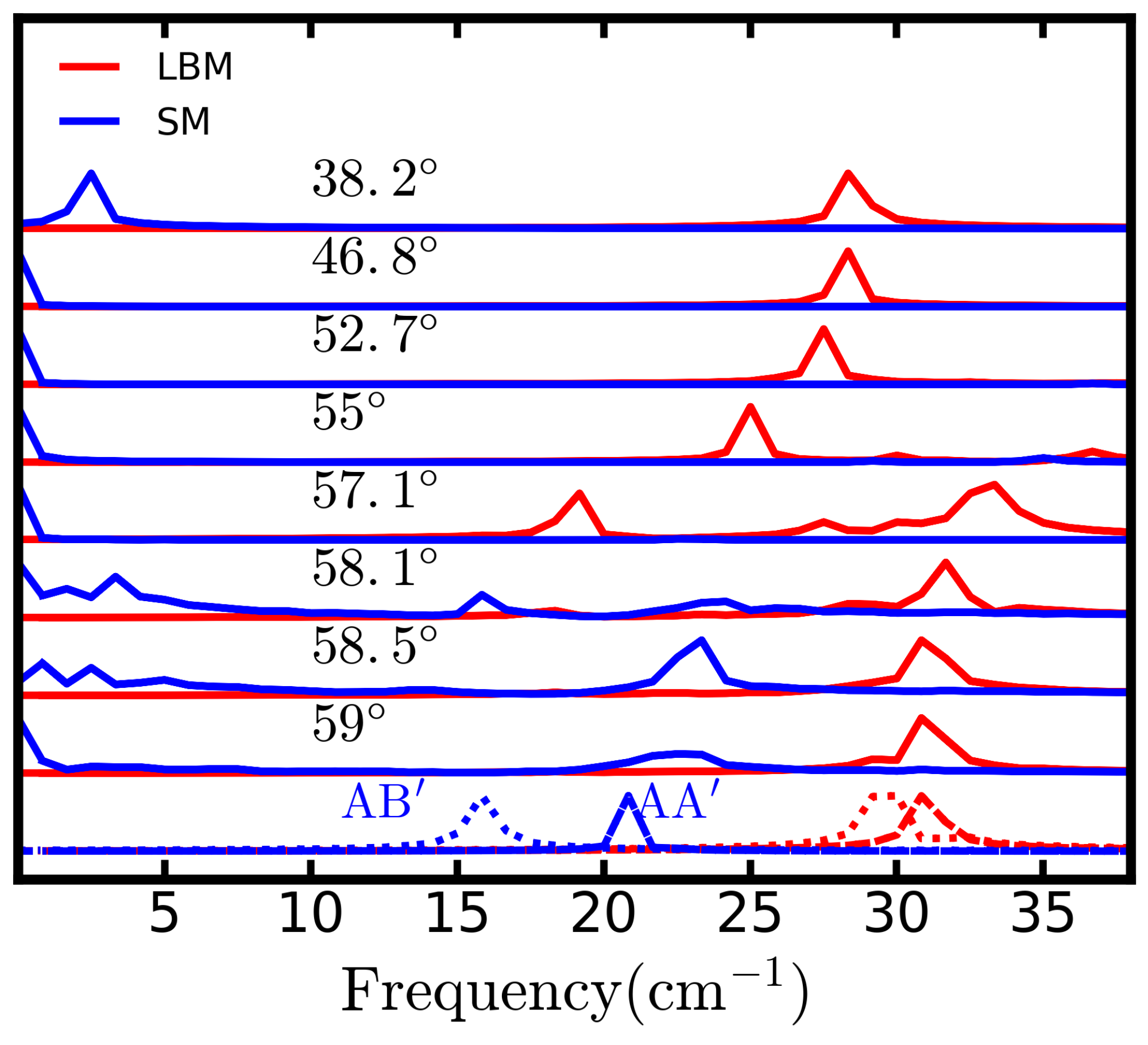}
		\end{subfigure}	
		\vspace*{-0.1cm}
		\caption{Evolution of low frequency vibrational modes with twist angles for twisted bilayer of $\mathrm{MoSe_{2}}$. The evolution of low frequency modes as $\theta \to 0^{\circ}$ are similar as $\theta \to 60^{\circ}$ and hence, not shown here.}
		\label{fig:mose2_example}
	\end{figure}

\textit{Other 2D materials: } Twisting one layer with respect to another in bilayers of 2D materials leads to the presence of multiple types of stacking in the MSL with different binding energies and stability, irrespective of the materials electronic properties. The intra-layer interaction in 2D materials are far stronger than the interlayer coupling. The combination of strong in-plane stiffness and weak variable interlayer coupling of twisted structures should produce similar behavior of low frequency vibrational modes in any 2D material. Thus, the existence of ultra-soft phason modes, the twist angle dependence of the corresponding eigenvectors and velocity are expected to be generic to any 2D materials. The twist angle dependence of the interlayer coupling is also revealed in the relaxed twisted structures. In order to illustrate, we compute the low frequency vibrational modes for twisted bilayer $\mathrm{MoSe_{2}}$ using mVACF at $T=300$ K (Fig.\ref{fig:mose2_example}). The trends are quite similar as in the case of $\mathrm{tBLMoS_{2}}$ justifying our conclusion. 

\textit{Raman Spectroscopy :} We have demonstrated the evolution of shear and layer breathing mode frequencies in twisted TMD bilayers. Irrespective of their Raman sensitivity, these modes are present in the twisted bilayer structures. For $\mathrm{BLMoS_{2}}$ (untwisted, most stable stacking), both the SM and LBM are found to be Raman active {\cite{Zhao_2013_nano}}. The observation of these modes using Raman spectroscopy depends on the Raman scattering intensity{\cite{Zhao_2013_nano}}. Since both the SM and LBM in the case of $\mathrm{BLMoS_{2}}$ are Raman active, we expect in the twisted structures SM and LBM with largest $p^\mathrm{SM}, p^{\mathrm{LBM}}$ will also be Raman active. However, this is speculative and explicit calculations of Raman intensity in twisted structures are unfeasible at present. Hence, we compare our results directly to Raman measurements in the case of $\mathrm{tBLMoSe_{2}}$ to show the usefulness of our calculations. The trends of the twist angle dependence of low frequency vibrational modes in $\mathrm{tBLMoSe_{2}}$ are in excellent agreement with experiment\cite{Puretzky_2016_acsnano}. Similar to $\mathrm{tBLMoS_{2}}$, we categorize the $\theta$ dependence of SM and LBM frequencies. In Region III, the high-frequency SM appears and LBM frequencies are similar to that of $\mathrm{BLMoSe_{2}}$, as predicted by our calculations. Multiple LBM and significant variation in LBM frequencies are also present in Region II. Both in Region I, II the SM frequencies are absent (because they are ultra-soft in nature) in Raman spectroscopy. However, two prominent features seem to be missing in the Raman experiment\cite{Puretzky_2016_acsnano} : (i) \textit{Monotonic} decrement of LBM frequencies in Region I (as in Fig.\ref{fig:lbm_regI}) and (ii) slight stiffening of the SM frequencies compared to $\mathrm{BLMoSe_{2}}$ (Region III. This might be due to the fact that, the experiment is carried out with chemical vapor deposition grown samples, which is known to exhibit lesser mobility and more disorder. We expect samples of better quality, such as mechanically exfoliated structures can show these missing features. 

	\begin{figure}
		\hspace*{-1cm}
		\begin{subfigure}[t]{0.45\textwidth}
			\vspace*{-0.3cm}
			\captionsetup{labelformat=empty}
			\caption{}
			\includegraphics[width=\textwidth]{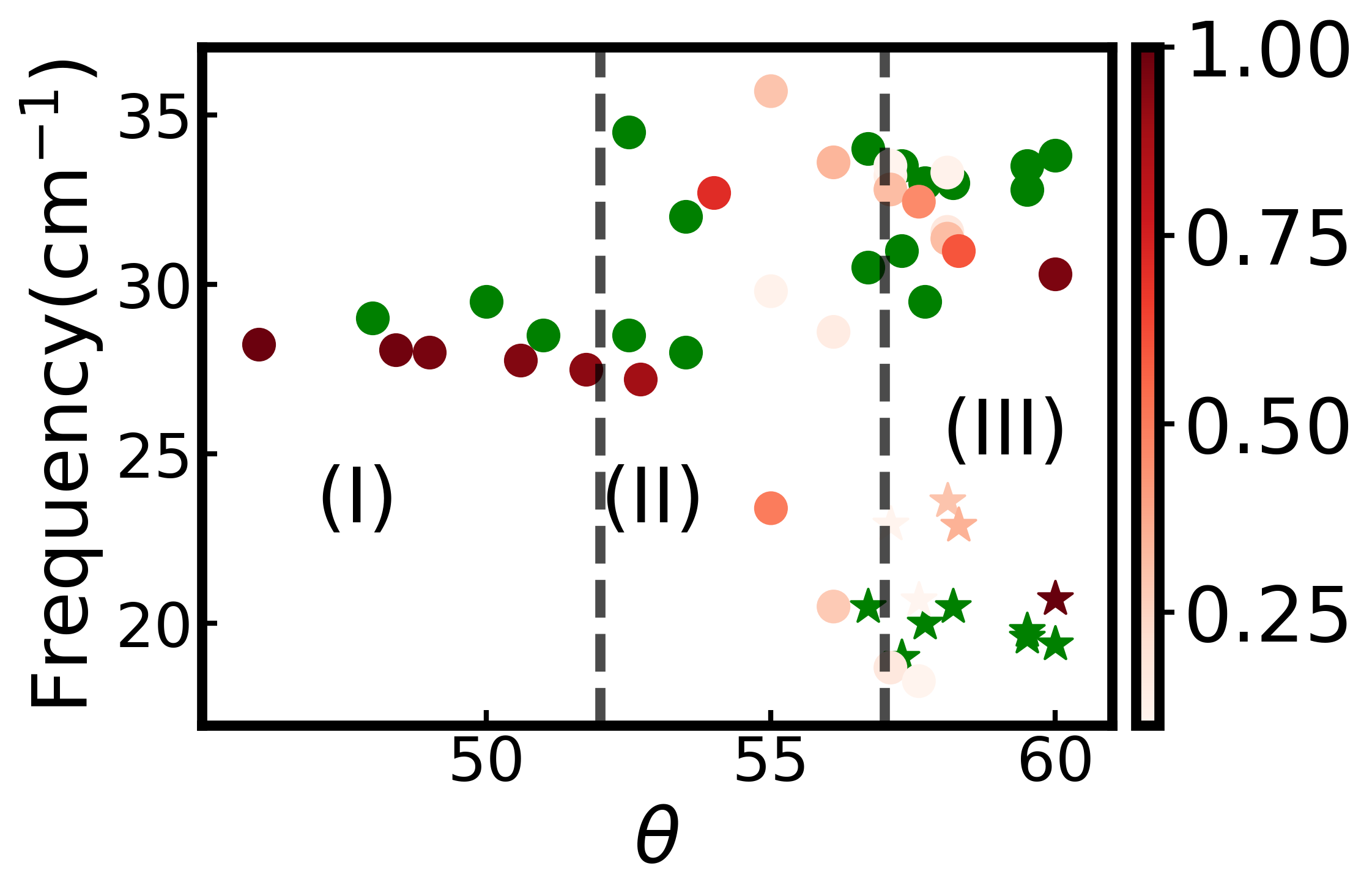}
		\end{subfigure}	
		\vspace*{-0.1cm}
		\caption{Comparison of the calculated twist angle dependence of low frequency vibrational modes and Raman measurements\cite{Puretzky_2016_acsnano} in $\mathrm{tBLMoSe_{2}}$. The experimental SMs (LBMs) are denoted with green star (circle), whereas the computed SMs (LBMs) with variable red color star (circle). The colorbar indicates $p^{\mathrm{SM}}, p^{\mathrm{LBM}}$. }
		\label{fig:mose2}
	\end{figure}

\textit{Brillouin-Mandelstam Spectroscopy : } Due to ultra-soft nature ($\lesssim 1 \ \mathrm{cm^{-1}}$ ) of the phason modes Raman spectroscopy can not be used to probe them. Brillouin-Mandelstam Spectroscopy (BMS) which can probe small frequencies (typically, $0.1-6 \ \mathrm{cm^{-1}}$), is often used in mineral physics and material science to probe acoustic modes\cite{Speziale_rmg_2014}. The phonon dispersion can also be mapped out by changing the incident light angle in BMS\cite{Kargar_natcom_2016}. Therefore, the sound speed along with elastic rigidity can be experimentally obtained. Using BMS spectroscopy our predictions of the existence of the ultra-soft modes can be verified. Also, by measuring the twist angle dependence of the dispersion of the acoustic and ultra-soft modes our results on velocity dependence of ultra-soft modes can be tested. Although, it is more likely to find incommensurate twisted structures in experiments and thus, the phason modes may become completely gapless. Still, we expect the phason velocity should be twist angle dependent as predicted here.

\begin{figure}[hbt!]
	\begin{subfigure}[H!]{0.46\textwidth}
		\captionsetup{labelformat=empty}
		\caption{}
		\label{fig:Cv}
		\includegraphics[width=\textwidth]{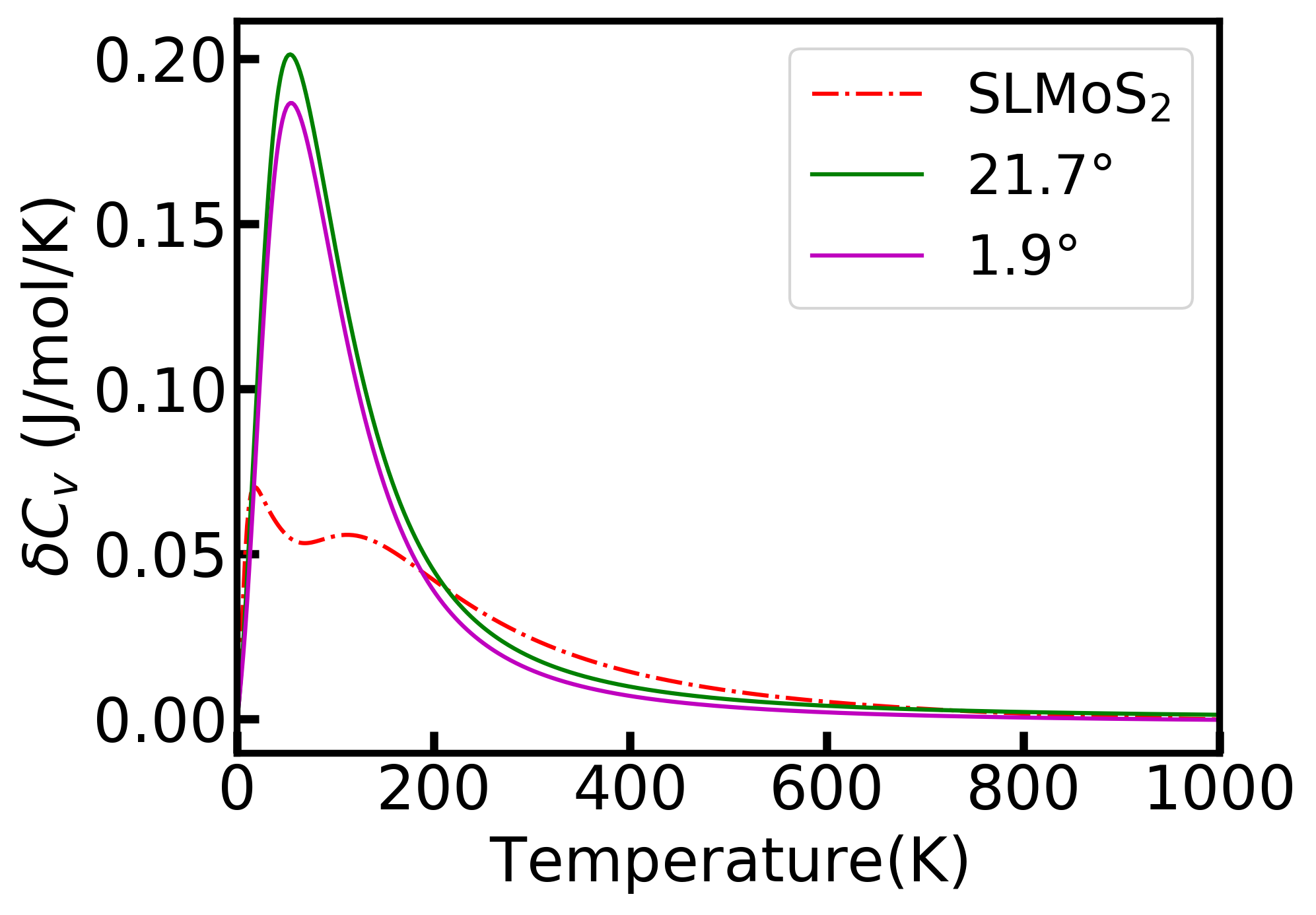}
	\end{subfigure} 
	\caption{ Twist angle dependence difference in specific heat (in units of J/mol/K) of $\mathrm{tBLMoS_{2}}$ with respect to $\mathrm{BLMoS_{2}}$ (untwisted).}
\end{figure}

\textit{More Twistnonics :} We have demonstrated the manipulation of low frequency vibrational modes with twist and related effects in frictional properties in transition metal dichalcogenide bilayers. Since phonon plays important role in determining many other material properties, the ability to control the phonon dispersion will facilitate engineering of those properties\cite{Balandin_jnn_2005}. Here, we outline a few of them : (i) The existence of ultra-soft modes can strongly modify the low temperature specific heat\cite{Hongyang_nanoscale_2014, Nika_apl_2014}, one of the key variable that dictates thermodynamic properties of a material. In order to illustrate this, we calculate the difference in specific heat of twisted structures with respect to $\mathrm{BLMoS_{2}}$. Clearly, there are significant changes in $\delta C_{v}$ for $T<200$ K. Corresponding single layer $\mathrm{MoS_{2}}$ value is also shown. (ii) Acoustic phonons are known to be the dominant heat carriers in insulating materials. The presence of ultra-soft phason modes can significantly modify the thermal conductivity of the twisted structures. Furthermore, as the group velocities of the ultra-soft phason modes can be tuned with twist angles, the thermal conductivity can also be engineered. (iii) The electron-phonon coupling for the ultra-soft modes can play important role in determining electrical resistivity. For example, in the case of twisted bilayer graphene electron-``acoustic" phonon coupling has been identified as the dominant source of it's high $T$ resistivity behavior\cite{Fengcheng_prb_2019}. The authors assumed the relative displacement of the two layers in twisted bilayer graphene as the additional ``acoustic" modes of the system with identical acoustic modes (LA, TA) dispersion. Although, this model captures the features of the resistivity qualitatively, the $D/\mathrm{v_{ph}}$ ($D$ is deformation potential for the electron-``acoustic" phonon coupling, $\mathrm{v_{ph}}$ is the phonon velocity) appearing in theoretical calculations seems to be off by a factor of 2-3 compared to experiment\cite{Fengcheng_prb_2019,Polshyn_arxiv_2019}. The assumption of the shear modes as the ``acoustic" modes is reasonable in the twisted structures, as they are ultra-soft in nature. However, the assumption that the dispersion of the ultra-soft modes and acoustic modes are identical is not accurate. Our calculations for the $\mathrm{tBLMoS_{2}}$ clearly shows the twist angle dependence of the velocity of the phason modes compared to that of the acoustic modes (by a factor of 2-3). This might provide an explanation for the missing factor in the case of tBLG. 

Since the SM and LBM are interlayer coupling dependent, by modifying the interlayer interaction strength (with external pressure, for example) the phonon spectra can be further engineered. Another important degree of freedom to tune phononic properties is the choice of materials, for instance, van der Waals heterostructures (stacking of two dissimilar 2D materials).

\section{\label{sec:level5}Conclusion}
Here, we have shown that the low frequency modes are extremely sensitive to twist in twisted TMD bilayer and can be used as a probe to determine twist angle. We also make falsifiable predictions of the presence of ultra-soft phason modes and the twist angle dependence of their eigenvectors, velocities. Our results indicate a twist angle dependent transition from superlubric to pinnied state. Our study provides the first step to \textit{twistnonics} in 2D materials.

\section{\label{sec:level5}Acknowledgement}
The authors thank Rahul Debnath, Shinjan Mandal, Sriram Ramaswamy, Sumilan Banerjee, Subroto Mukerjee, Arindam Ghosh and Rahul Pandit for useful discussions and thank the Supercomputer Education and Research Center (SERC) at IISc for providing computational resources.

\textit{Note Added:} While submitting the revised version we become aware of two recent studies, carried out within an elastic continuum approximation, reporting the existence of similar ultra-soft phonon modes in tBLG\cite{koshino2019moire, ochoa2019moire}.

\newpage


%

\end{document}